\documentclass[aps,prl,showpacs,superscriptaddress,groupedaddress,nofootinbib]{revtex4-2}  
\usepackage{cancel}
\usepackage{graphicx} 
\usepackage[T1]{fontenc}
\usepackage{lmodern} 
\usepackage{graphicx}  
\usepackage{dcolumn}   
\usepackage{bm}        
\usepackage{amssymb}   
\usepackage{amsmath}
\usepackage{bbold}
\usepackage{array}
\usepackage{float}
\usepackage{makecell}
\usepackage{braket}
\usepackage{longtable}
\usepackage{supertabular,booktabs}
\usepackage{soul}

\usepackage{titlesec} 
\usepackage[colorlinks,citecolor=blue,urlcolor=blue,hypertexnames=true]{hyperref}
\usepackage{subfigure}

\usepackage[usenames,dvipsnames,svgnames,table]{xcolor} 

\renewcommand{\arraystretch}{2}

\newcommand{\be}{\begin{equation}}
\newcommand{\ee}{\end{equation}}
\newcommand{\bea}{\begin{eqnarray}}
\newcommand{\eea}{\end{eqnarray}}
\newcommand{\bml}{\begin{subequations}}
\newcommand{\eml}{\end{subequations}}
\newcommand{\bfig}{\begin{figure}}
\newcommand{\efig}{\end{figure}}

\newcommand{\slashed}{\hspace{-1.1ex}/}

\newcommand{\bmat}{\begin{pmatrix}}
\newcommand{\emat}{\end{pmatrix}}
\usepackage{graphicx, slashed,booktabs, color, multirow, float,
amsfonts, bbold, mathtools, sidecap, tikz, bm,enumitem}
\usepackage{multirow}
\usepackage{bbding}
\usepackage{titlesec}
\usepackage{hyperref}
\usepackage{wasysym}
\usepackage{amssymb}
\usepackage{pifont}

\renewcommand{\leq}{\leqslant}

\usepackage[dvipsnames, usenames]{xcolor}

\definecolor{linkcolor}{rgb}{0.55, 0.13, .32}

\definecolor{oucrimsonred}{rgb}{0.6, 0.0, 0.0}
\definecolor{persianblue}{rgb}{0.11, 0.22, 0.73}
\definecolor{forestgreen}{rgb}{0.13,0.35,0.13}
\definecolor{lightgray}{rgb}{0.83, 0.83, 0.83}
 \hypersetup{colorlinks, citecolor=oucrimsonred, linkcolor=persianblue, urlcolor=oucrimsonred}
\definecolor{cornellred}{rgb}{0.7, 0.11, 0.11}
\definecolor{navyblue}{rgb}{0.0, 0.0, 0.5}
\definecolor{amethyst}{rgb}{0.6, 0.4, 0.8}
\definecolor{yellow}{rgb}{1.0, 1.0, 0.0}
\definecolor{firebrick}{rgb}{0.7, 0.13, 0.13}
\definecolor{tangerineyellow}{rgb}{1.0, 0.8, 0.0}
\definecolor{deepfuchsia}{rgb}{0.76, 0.33, 0.76}
\definecolor{amber}{rgb}{1.0, 0.75, 0.0}
\definecolor{VioletRed4}{rgb}{0.55, 0.13, .32}
\definecolor{indiagreen}{rgb}{0.07, 0.53, 0.03}
\definecolor{VioletRed4}{rgb}{0.55, 0.13, .32}

\usepackage{hyperref}
\usepackage{graphics, appendix, afterpage, makecell} 
\usepackage{bbold}
\usepackage{tikz}
\usepackage{adjustbox}

\usepackage{tcolorbox}

\definecolor{oucrimsonred}{rgb}{0.6, 0.0, 0.0}
\definecolor{persianblue}{rgb}{0.11, 0.22, 0.73}
\definecolor{forestgreen}{rgb}{0.13,0.35,0.13}
\definecolor{lightgray}{rgb}{0.83, 0.83, 0.83}
 \hypersetup{colorlinks, citecolor=oucrimsonred, linkcolor=persianblue, urlcolor=oucrimsonred}
\definecolor{cornellred}{rgb}{0.7, 0.11, 0.11}
\definecolor{navyblue}{rgb}{0.0, 0.0, 0.5}
\definecolor{amethyst}{rgb}{0.6, 0.4, 0.8}
\definecolor{yellow}{rgb}{1.0, 1.0, 0.0}
\definecolor{firebrick}{rgb}{0.7, 0.13, 0.13}
\definecolor{tangerineyellow}{rgb}{1.0, 0.8, 0.0}
\definecolor{deepfuchsia}{rgb}{0.76, 0.33, 0.76}
\definecolor{amber}{rgb}{1.0, 0.75, 0.0}
\definecolor{VioletRed4}{rgb}{0.55, 0.13, .32}
\definecolor{indiagreen}{rgb}{0.07, 0.53, 0.03}
\definecolor{VioletRed4}{rgb}{0.55, 0.13, .32}

\definecolor{oucrimsonred}{rgb}{0.6, 0.0, 0.0}
\newcommand\vertarrowbox[3][6ex]{%
  \begin{array}[t]{@{}c@{}} #2 \\
  \left\uparrow\vcenter{\hrule height #1}\right.\kern-\nulldelimiterspace\\
  \makebox[0pt]{\scriptsize#3}
  \end{array}%
}

\definecolor{mtcolor}{rgb}{.8,.3,.1}

\definecolor{violachiaro}{rgb}{1,0.6,1}

\definecolor{gbcolor}{rgb}{.43,.22,.12}
 
\definecolor{gbcolor2}{rgb}{.9,.2,.6}
\definecolor{gbcolor3}{rgb}{.3,.2,.6}

\definecolor{verdechiaro}{rgb}{0.6,1,0.6}
\definecolor{giallochiaro}{rgb}{1,1,0.6}
\definecolor{bluscuro}{rgb}{0.15, 0.2, 0.9}
\definecolor{verdes}{rgb}{0.1, 0.5, 0.1}%
\definecolor{tangerineyellow}{rgb}{1.0, 0.8, 0.0}
\definecolor{smokyblack}{rgb}{0.06, 0.05, 0.03}

\definecolor{americanrose}{rgb}{1.0, 0.01, 0.24}
\definecolor{cobalt}{rgb}{0.0, 0.28, 0.67}
\definecolor{brandeisblue}{rgb}{0.0, 0.44, 1.0}
\definecolor{mycolor}{rgb}{0.0, 0.0, 0.5}
\definecolor{oxfordblue}{rgb}{0.0, 0.13, 0.28}
\definecolor{azure}{rgb}{0.0, 0.5, 1.0}
\definecolor{turquoiseblue}{rgb}{0.0, 1.0, 0.94}
\newtcolorbox{mynewbox}[1]{colback=white!5!white,colframe=azure!75!black,fonttitle=\bfseries,title=#1}
\newtcolorbox{mybox}{colback=mycolor!5!white,colframe=azure!75!black}
\newtcolorbox{mynamedbox}[1]{colback=mycolor!5!white,colframe=azure!75!black,title=#1}
\definecolor{venetianred}{rgb}{0.78, 0.03, 0.08}
\newtcolorbox{mynamedbox1}[1]{colback=venetianred!5!white,colframe=venetianred!80!black,title=#1}
\newtcolorbox{mynamedbox2}[1]{colback=azure!5!white,colframe=azure!80!black,title=#1}

\definecolor{rossocorsa}{rgb}{0.83, 0.0, 0.0}

\tikzset{->-/.style={decoration={
  markings,
  mark=at position #1 with {\arrow{>}}},postaction={decorate}}}
\tikzset{-<-/.style={decoration={
  markings,
  mark=at position #1 with {\arrow{<}}},postaction={decorate}}} 

\def\be{\begin{equation}}
\def\ee{\end{equation}}
\def\ba{\begin{eqnarray}}
\def\ea{\end{eqnarray}}

\def\L*{{\cal L}_*}
\def\L{\mathcal{L}}
\def\({\left(}
\def\){\right)}

\def\<{\langle}
\def\>{\rangle}

 \def\neq {\not\equiv}

\def\cs2{c_{s}^{2}}

 \def\be   {\begin{equation}}   \def\ee   {\end{equation}}
 \def\ba   {\begin{array}}      \def\ea   {\end{array}}
 \def\bea  {\begin{eqnarray}}   \def\eea  {\end{eqnarray}}
 \def\bean {\begin{eqnarray*}}  \def\eean {\end{eqnarray*}}

\titleclass{\subsubsubsection}{straight}[\subsection]

\newcounter{subsubsubsection}[subsubsection]
\renewcommand\thesubsubsubsection{\thesubsubsection.\arabic{subsubsubsection}}

\titleformat{\subsubsubsection}
  {\normalfont\normalsize\bfseries}{\thesubsubsubsection}{1em}{}
\titlespacing*{\subsubsubsection}
{0pt}{3.25ex plus 1ex minus .2ex}{1.5ex plus .2ex}

\makeatletter
\renewcommand\paragraph{\@startsection{paragraph}{5}{\z@}%
  {3.25ex \@plus1ex \@minus.2ex}%
  {-1em}%
  {\normalfont\normalsize\bfseries}}
\renewcommand\subparagraph{\@startsection{subparagraph}{6}{\parindent}%
  {3.25ex \@plus1ex \@minus .2ex}%
  {-1em}%
  {\normalfont\normalsize\bfseries}}
\def\toclevel@subsubsubsection{4}
\def\toclevel@paragraph{5}
\def\toclevel@paragraph{6}
\def\l@subsubsubsection{\@dottedtocline{4}{7em}{4em}}
\def\l@paragraph{\@dottedtocline{5}{10em}{5em}}
\def\l@subparagraph{\@dottedtocline{6}{14em}{6em}}
\makeatother

\usepackage{titlesec} 
\usepackage[colorlinks,citecolor=blue,urlcolor=blue,hypertexnames=true]{hyperref}
\usepackage{subfigure}

\usepackage[usenames,dvipsnames,svgnames,table]{xcolor}

\usepackage{tikzsymbols}
\usepackage{natbib}
\usepackage{float}

\usepackage{tikz,xcolor,hyperref}

\usepackage{slashed}

\definecolor{lime}{HTML}{A6CE39}
\DeclareRobustCommand{\orcidicon}{
	\begin{tikzpicture}
	\draw[lime, fill=lime] (0,0) 
	circle [radius=0.2] 
	node[white] {{\fontfamily{qag}\selectfont \tiny ID}};
	\draw[white, fill=white] (-0.0625,0.095) 
	circle [radius=0.007];
	\end{tikzpicture}
	\hspace{-2mm}
}

\usepackage[usenames,dvipsnames,svgnames,table]{xcolor}  
\usepackage{tikzsymbols}
\usepackage{natbib}
\usepackage{float}

\usepackage{tikz,xcolor,hyperref}

\usepackage{slashed}

\definecolor{lime}{HTML}{A6CE39}
\DeclareRobustCommand{\orcidicon}{
	\begin{tikzpicture}
	\draw[lime, fill=lime] (0,0) 
	circle [radius=0.2] 
	node[white] {{\fontfamily{qag}\selectfont \tiny ID}};
	\draw[white, fill=white] (-0.0625,0.095) 
	circle [radius=0.007];
	\end{tikzpicture}
	\hspace{-2mm}
}

\foreach \x in {A, ..., Z}{\expandafter\xdef\csname orcid\x\endcsname{\noexpand\href{https://orcid.org/\csname orcidauthor\x\endcsname}
			{\noexpand\orcidicon}}
}
 
\usepackage{bbold}
\usepackage{tikz}
\usepackage{adjustbox}
\usepackage{tcolorbox}
\usepackage{enumitem}
\usepackage{amsfonts}

\setlist[itemize,1]{label=$\times$}
\setlist[itemize,2]{label=$\checkmark$}
\setlist[itemize,3]{label=$\diamond$}
\setlist[itemize,4]{label=$\bullet$}

\begin{document}
\title{\Large \textcolor{Sepia}{What new physics can we extract from inflation using the ACT DR6 and DESI DR2 Observations?
}}
\author{\large Sayantan Choudhury\orcidA{}${}^{1,2}$}
\email{sayanphysicsisi@gmail.com, \\
sayantan_ccsp@sgtuniversity.org,\\
sayantan.choudhury@nanograv.org (Corresponding author)} 
\author{\large Gulnur Bauyrzhan \orcidC{}${}^{3}$}
\email{baurzhan.g.b@gmail.com} 
\author{\large Swapnil Kumar Singh\orcidF{}${}^{4}$}
\email{swapnil.me21@bmsce.ac.in, swapnilsingh.ph@gmail.com}
\author{\large Koblandy Yerzhanov\orcidB{}${}^{3}$}
\email{yerzhanovkk@gmail.com}

\affiliation{${}^{1}$ Centre for Cosmology and Science Popularization (CCSP), SGT University, Gurugram, Delhi- NCR, Haryana- 122505, India.}

\affiliation{ ${}^{2}$Institute of Theoretical Physics, Faculty of Physics,\\
University of Warsaw, ul. Pasteura 5, 02-093 Warsaw, Poland,}
\affiliation{ ${}^{3}$Center for Theoretical Physics,  L.N. Gumilyov Eurasian National University, Astana 010008, Kazakhstan,}
\affiliation{ ${}^{4}$B.M.S. College of Engineering, 
    Bangalore, Karnataka, 560019, India.}







\begin{abstract}
We present a comprehensive analysis of inflationary models in light of projected sensitivities from forthcoming CMB and gravitational wave experiments, incorporating data from recent ACT DR6, DESI DR2, CMB-S4, LiteBIRD, and SPHEREx. Focusing on precise predictions in the $(n_s, \alpha_s, \beta_s)$ parameter space, we evaluate a broad class of inflationary scenarios---including canonical single-field models, non-minimally coupled theories, and string-inspired constructions such as Starobinsky, Higgs, Hilltop, $\alpha$-attractors, and D-brane models. Our results show that next-generation observations will sharply constrain the scale dependence of the scalar power spectrum, elevating $\alpha_s$ and $\beta_s$ as key discriminants between large-field and small-field dynamics. Strikingly, several widely studied models---such as quartic Hilltop inflation and specific DBI variants---are forecast to be excluded at high significance. We further demonstrate that the combined measurement of $\beta_s$ and the field excursion $\Delta\phi$ offers a novel diagnostic of kinetic structure and UV sensitivity. These findings underscore the power of upcoming precision cosmology to probe the microphysical origin of inflation and decisively test broad classes of theoretical models.

\end{abstract}

\maketitle

\section{Introduction}

A major paradigm shift in the history of theoretical physics was brought about by the development of the framework of cosmological inflation \cite{Choudhury:2024aji,Baumann:2009ds,Senatore:2016aui,Kallosh:2025ijd,Odintsov_2023}. Beyond resolving conceptual shortcomings of the standard hot Big Bang cosmology---most notably the horizon, flatness, and monopole problems---inflation provides a dynamical mechanism for generating primordial quantum fluctuations that seed the observed large-scale structure of the Universe. This remarkable connection between quantum field theory in curved spacetime and cosmological observations has established inflation as a cornerstone of modern cosmology, effectively bridging high-energy physics and observational astrophysics.

Following its development, cosmology has evolved into a precision science, enabling increasingly stringent tests of fundamental physics. In this context, a natural and timely question arises: beyond the extensive body of work on inflationary model building and phenomenology, is it possible to extract qualitatively new physical insights from existing frameworks in light of current observational precision? More specifically, given the latest cosmological datasets, can inflation provide additional information about high-energy physics that goes beyond conventional observables?

In this work, we address these questions using the most recent observational inputs, particularly the latest data releases from the Atacama Cosmology Telescope (ACT) \cite{ACT:2025fju,ACT:2025tim} and the Dark Energy Spectroscopic Instrument (DESI) DR2 \cite{DESI:2024mwx,DESI:2024uvr}. These datasets, when combined with earlier measurements, lead to significant refinements in the constraints on inflationary observables, thereby opening new directions for model discrimination and phenomenological exploration.

The recent ACT data release provides strong support for the inflationary paradigm. However, when combined with DESI DR2 observations, the inferred constraints on key parameters---particularly the scalar spectral index $n_s$ and the tensor-to-scalar ratio $r$---undergo notable shifts. A number of recent studies have explored the implications of these combined datasets for inflationary physics (see, e.g., \cite{Kallosh:2025rni,Dioguardi:2025vci,Brahma:2025dio,Gialamas:2025kef,Salvio:2025izr,Dioguardi:2025mpp,Gao:2025onc,Drees:2025ngb,Yin:2025rrs,Liu:2025qca,Gialamas:2025ofz,Byrnes:2025kit,Addazi:2025qra,Yi:2025dms,Peng:2025bws,Frolovsky:2025iao,pallis2025kineticallymodifiedpalatiniinflation,aoki2025higgsmodularinflation,mondal2025constrainingreheatingtemperatureinflatonsm,wolf2025inflationaryattractorsradiativecorrections}), highlighting the sensitivity of inflationary predictions to updated observational inputs.

For reference, the Planck 2018 results constrain the scalar spectral index to $n_s = 0.9651 \pm 0.0044$ \cite{Planck:2018vyg}. Subsequent analyses have indicated a mild upward shift, for example $n_s = 0.9683 \pm 0.0040$ \cite{Efstathiou:2019mdh}. The combination of Planck and ACT data (P-ACT) further increases this value to $n_s = 0.9709 \pm 0.0038$, while the inclusion of DESI data (P-ACT-LB) yields $n_s = 0.9743 \pm 0.0034$. This represents an approximately $2\sigma$ shift relative to the original Planck result, and has important implications for inflationary model selection.

In particular, plateau-type models such as the Starobinsky model \cite{Starobinsky:1980te}, which are in excellent agreement with Planck-only constraints, can become mildly disfavored at the $\gtrsim 2\sigma$ level when confronted with the higher central value of $n_s$ inferred from the P-ACT-LB dataset. However, this conclusion is not definitive and depends sensitively on the dataset combination adopted. Notably, the recent SPT/Planck analysis reports a lower value, $n_s = 0.9647 \pm 0.0037$ \cite{SPT-3G:2024atg}, which is fully consistent with the Planck baseline and restores compatibility with plateau-type models.

This $\sim 2\sigma$ discrepancy between different dataset combinations motivates a cautious interpretation of current constraints. If the P-ACT-LB central value is adopted, models capable of producing relatively larger $n_s$, such as certain hilltop-type scenarios, become comparatively more favored, while Starobinsky-like plateau models move toward the lower edge of the preferred region. Conversely, if the SPT/Planck central value is adopted, the Starobinsky and Higgs-like plateau predictions remain well aligned with the data, while models requiring higher $n_s$ lose some of their comparative advantage. For this reason, our conclusions are phrased in terms of dataset-dependent viability rather than absolute exclusion.

It is useful to make this dataset dependence explicit. If the SPT/Planck central value is used as the reference point, plateau models such as Starobinsky and Higgs inflation move back toward the center of the preferred region, whereas models whose comparative advantage relies on larger values of $n_s$ become less favored. If instead the P-ACT-LB central value is treated as the relevant benchmark, the same plateau models are not excluded but sit closer to the lower edge of the preferred range, while hilltop-like scenarios can appear comparatively better aligned. The model ranking reported below should therefore be understood as conditional on the adopted data combination rather than as an absolute hierarchy of theoretical viability.

Motivated by these considerations, in this work we perform a comprehensive numerical analysis of several well-motivated single-field inflationary models with non-minimal gravitational couplings. In addition to the standard observables $(n_s, r)$, we place particular emphasis on higher-order spectral parameters, namely the running $\alpha_s$ and the running of the running $\beta_s$, as potential discriminants for future observations. We also investigate reheating dynamics, inflationary energy scales, field excursion limits, and implications for primordial gravitational waves within a unified computational framework.

The remainder of this paper is organized as follows. In Section~\ref{sec:Important}, we outline the key physical questions that motivate our analysis, including the role of non-minimal couplings, reheating dynamics, higher-order spectral observables, field-excursion constraints, and primordial gravitational waves.

Section~\ref{sec:methodology} presents the numerical framework used to compute inflationary observables, including the conformal transformation from the Jordan frame to the Einstein frame, the evaluation of slow-roll parameters, and the reconstruction of observables from the canonically normalized potential.

In Section~\ref{sec:models}, we introduce the general class of non-minimally coupled single-field inflationary models, while Section~\ref{sec:models_under_study} specifies the benchmark potentials studied in this work, namely the Starobinsky model, Higgs inflation, the $\alpha$-attractor T-model, quartic Hilltop inflation, and D-brane inflation.

Section~\ref{sec:results} presents the main numerical results, including predictions for inflationary observables, reheating temperatures, inflationary Hubble scales, and field excursions across the parameter space of each model.

In Section~\ref{sec:discriminating}, we analyze the discriminating power of current and future CMB observations in the $(n_s,\alpha_s)$ plane, incorporating projected sensitivities from experiments such as CMB-S4 and SPHEREx.

Sections~\ref{sec:running_and_beta} and~\ref{sec:field_excursion} are devoted to higher-order spectral observables and field-excursion constraints, respectively, highlighting their role in probing physics beyond the leading-order $(n_s,r)$ description.

In Section~\ref{sec:gw}, we study primordial gravitational waves, including predictions for the tensor spectrum and the stochastic gravitational-wave background, taking into account the effect of reheating assumptions.

Section~\ref{sec:model_comparison} provides a comprehensive comparison of all models, summarizing their predictions for key observables such as $n_s$, $r$, $\Delta\varphi$, $H_{\rm inf}$, and $T_{\rm reh}$.

Finally, Section~\ref{sec:conc} summarizes our main conclusions and discusses future directions.

\section{Key Physical Questions}
\label{sec:Important}

Some of the key questions that motivate our analysis from the perspective of extracting new physical information from the inflationary paradigm are listed below:
\begin{enumerate}
    \item In the usual circumstances for inflation, one uses the canonical or non-canonical scalar field minimally coupled with the Einstein-Hilbert term. However, such frameworks are often too restrictive and, in some cases, are increasingly constrained by recent observational data as mentioned above. In such a situation, one needs a modification in the corresponding defining theory of inflation. Among various possibilities, one of the most promising scenarios is to incorporate a non-minimal coupling $\xi$ with inflation and the representative Einstein-Hilbert term, i.e., by incorporating a contribution after the modification $\sqrt{-g}\left(1+\xi f(\phi)\right) R$ \footnote{Recently in ref. \cite{Kallosh:2025rni}, the authors have already pointed out that with the choice of the non-minimal coupling $\xi=1$ and the function $f(\phi)=\phi/M_{\rm pl}$, which actually gives rise to $\sqrt{-g}\left(1+\phi/M_{\rm pl}\right) R$ along with quadratic effective potential, $V(\phi)=\frac{1}{2}m^2\phi^2$ is strongly supported by the constraints from the present observational probes.}. Here $f(\phi)$ is a general function of the inflaton field $\phi$. For model-building purposes, one can choose the simplest power-law form of the function $f(\phi)=(\phi/M_{\rm pl})^n$, where, $n=1,2,3,\cdots$ can be considered. See refs. \cite{Kallosh:2025rni} where such possibilities along with the non-minimal coupling $\xi$ have been considered previously. As an example, a specific template of $f(\phi)=(\phi/M_{\rm pl})^2$ with Higgs effective potential is usually considered to describe the Higgs inflationary paradigm \cite{Bezrukov:2007ep,Rubio:2018ogq}. Recently in ref. \cite{Kallosh:2025rni} to confront the present observational probes, the author has chosen a specific case of the effective potential $V(\phi)=\lambda^2f^2(\phi)$ with $f(\phi)=(\phi/M_{\rm pl})^n$, which means that once the functional form of the non-minimal coupling function $f(\phi)$ is fixed that automatically fixes the structure of the inflationary effective potential, here particularly in even power-law format. Though in ref. \cite{Kallosh:2025rni}, the authors have claimed that with this type of choice, one can cover a large class of power-law type of inflationary models, in this paper we relax this phenomenologically motivated assumption. Instead of choosing a connecting relationship between the non-minimal coupling function $f(\phi)$ and the inflationary effective potential $V(\phi)$ in this paper we consider any arbitrary form of the effective potential $V(\phi)$ that can be derived from various UV-complete high-energy physics scenarios. The advantage of relaxing this assumption is that it allows us to cover a large class of inflationary models, a broader class of models that have not been covered recently in ref. \cite{Kallosh:2025rni}. Additionally, it is important to note that, $g={\rm det}(g_{\mu\nu})$ is the determinant of the background metric $g_{\mu\nu}$, which describes the classical geometry of $3+1$-dimensional FLRW spatially flat space-time with a quasi-de Sitter solution in the Jordan frame. In this frame, the representative model studied in this paper is described by,
\begin{eqnarray}
    \frac{1}{\sqrt{-g}}{\cal L}=\left[\frac{M^2_{\rm pl}}{2}\left(1+\xi f(\phi)\right)R-\frac{1}{2}\left(\partial\phi\right)^2-V(\phi)\right].
\end{eqnarray}
Here, $M_{\rm pl}=M_p/\sqrt{8\pi}=2.43\times 10^{18}$ GeV is the reduced Planck mass, and $M_p=10^{19}$ GeV is the actual Planck mass.

For clarity, in the general discussion of this section we denote the inflaton by $\phi$, while in the numerical analysis we reserve $\varphi$ for the canonically normalized Einstein frame field.
The first term describes the modification of the Einstein-Hilbert term after the inclusion of non-minimal coupling through the parameter $\xi$ and function $f(\phi)$. The second and third terms describe the canonical kinetic term (where we define, $(\partial\phi)^2=g^{\mu\nu}\partial_{\mu}\phi\partial_{\nu}\phi$) and the effective potential of the inflation in the Jordan frame \footnote{One can further consider the more generalized version of our proposed model for further analysis by incorporating both non-minimal coupling with the Einstein-Hilbert term and non-canonical contribution in the inflaton sector in the Jordan frame, which is described by,
   \begin{eqnarray}
    \frac{1}{\sqrt{-g}}{\cal L}=\left[\frac{M^2_{\rm pl}}{2}\left(1+\xi f(\phi)\right)R+P(X,\phi)\right],
\end{eqnarray} 
where $X=-1/2(\partial\phi)^2$ represents the kinetic term. Here $P(X,\phi)$ is a general functional form that can incorporate some other classes of inflationary models, such as DBI, Tachyon, and GTachyon models, studied previously without incorporating the non-minimal coupling parameter $\xi$ (i.e., for these models other studies have been done with $\xi=0$.). These studies introduce an effective sound speed, $c_s$ which is an additional parameter. See refs. \cite{Choudhury:2015hvr,Choudhury:2017glj} for more details on this issue. Here with $\xi=0$ and canonical model where $P(X,\phi)=X-V(\phi)$ one can get back the known result $c_s=1$. It is further expected that the inclusion of non-minimal coupling will further modify and complicate the expression for the effective sound speed $c_s$. After the conversion of the action in the Einstein frame, it is further expected that the functional $P(X,\phi)$ will be modified as $P(X,\phi)/\left(1+\xi f(\phi)\right)^2$ and the Einstein-Hilbert term will be free from the non-minimal coupling. However, to make our analysis simpler, we have not considered this particular possibility in this paper, which we may address in due course.} \footnote{Apart from the proposed model in this paper, one can further consider the possibility of having a physical scenario where the non-minimal coupling is introduced in front of the Gauss-Bonnet higher derivative terms (which is commonly known as Gauss-Bonnet inflation) in the Jordan frame, which is described by,
\begin{eqnarray}
    \frac{1}{\sqrt{-g}}{\cal L}=\left[\frac{M^2_{\rm pl}}{2}R+\xi f(\phi)R^2_{\rm GB}+X-V(\phi)\right],
\end{eqnarray}
where $R^2_{\rm GB}=\left(R^{\mu\nu\alpha\beta}R_{\mu\nu\alpha\beta}-4R^{\mu\nu}R_{\mu\nu}+R^2\right)$ is the Gauss-Bonnet term. In $3+1$-dimensional space-time, when the non-minimal coupling parameter $\xi=1$ and the function $f(\phi)=1$ it turns out to be topological terms, they can be expressed as a total divergence contribution and hence are redundant for the study of inflation. So to study inflation, such non-minimal coupling with the Gauss-Bonnet term having some functional structure of $f(\phi)$ (i.e. $f(\phi)\neq 1$) is necessarily required. See refs. \cite{Kanti:2015pda,Mudrunka:2023wxy,Nojiri:2023mvi,Fernandes:2022zrq, odintsov2025gw170817viableeinsteingaussbonnetinflation, Odintsov_2018, odintsov2020rectifyingeinsteingaussbonnetinflationview} for more details on this issue. The rest of the part in the above action represents the usual canonical theory of inflation, which can be further modified by introducing general functional $P(X,\phi)$ to study the larger class of inflationary models in a non-canonical regime. However, to make our analysis simpler, we have not considered this particular possibility in this paper, which we may address in due course.} Further performing a conformal transformation from the Jordan to the Einstein frame through the metric, $\hat{g}_{\mu\nu}=\Omega^2 g_{\mu\nu}$ one can translate our proposed model in a simpler form, which is given by:
\begin{eqnarray}
    \frac{1}{\sqrt{-\hat{g}}}\hat{\cal L}=\left[\frac{M^2_{\rm pl}}{2}\hat{R}-\frac{1}{2}\left(\hat{\partial}\phi\right)^2-\hat{V}(\phi)\right].
\end{eqnarray}
where in the Einstein frame the first term represents the Einstein-Hilbert contribution, the second term is the canonical kinetic term of the inflaton, and the new effective potential is given by the following expression:
\begin{eqnarray}
    \hat V(\phi)=\frac{V(\phi)}{\left(1+\dfrac{\xi f(\phi)}{M_{\rm pl}^2}\right)^2}.
\end{eqnarray} 
See refs. \cite{Choudhury:2015zlc,Choudhury:2017cos} for more details on this transformation. A natural question then arises: is it possible to constrain the value of the non-minimal coupling parameter by imposing the constraints on the tensor-to-scalar ratio $r$ and spectral index $n_s$ from the recent ACT and DESI DR2 data release (along with considering other combined constraints from observation)? In this work, we address this question for a broad class of canonical inflationary models, with the aim of extracting new physical information from current observational constraints.

    \item So far, we have discussed the usual cold canonical inflationary paradigm, including non-minimal coupling to the Einstein gravity. However, there might be a possibility of studying a warm inflationary paradigm in the light of recent observations. Recently in ref. \cite{Berera:2025vsu} the authors have claimed that the early universe may admit an ACT-motivated warm-inflation interpretation. Although we do not assess the validity of this specific interpretation here, it provides an interesting possibility for future work on inflationary models supported by the warm-inflation paradigm. Explicitly speaking, we have not studied this possibility in this paper but leave this possibility for future investigation. 

    \item Using the new ACT and DESI DR2 data, we constrain the tensor-to-scalar ratio $r$ for the inflationary models studied in this paper. In the slow-roll regime, this information allows us to estimate the corresponding inflationary energy scale through the corresponding scale of inflation by using the following equation:
    \begin{eqnarray}
        V_{\rm inf}=3M^2_{\rm pl}H^2_{\rm inf}=\left(1.96\times 10^{16}{\rm GeV}\right)^4\times \left(\frac{r}{0.12}\right).
    \end{eqnarray}
Using the above equation, one can fix both $V_{\rm inf}$ and $H_{\rm inf}$ for a given value of $r$. Once the scale of inflation is fixed through the value of $r$ at the CMB scale (at $k=k_{\rm cmb}$) one can place useful constraints on possible UV-complete scenarios and viable high-energy physics models responsible for inflation. Finding new information about physics obviously plays a significant role in the present study. 

    \item It is also possible to further constrain the reheating temperature once the scale of inflation is fixed through the tensor-to-scalar ratio $r$ using the new ACT and DESI DR2 data. This can be estimated using the following expression:
    \begin{eqnarray}
        T_{\rm reh}&=&\left(\frac{90}{\pi^2g_{\star}}\right)^{1/4}\sqrt{H_{\rm inf}M_{\rm pl}}\nonumber\\
        &=&\left(\frac{90}{\pi^2g_{\star}}\right)^{1/4}\left(\frac{r}{0.12}\right)^{1/4}\frac{\left(1.96\times 10^{16}{\rm GeV}\right)}{\sqrt{3}}.
    \end{eqnarray}
Here $g_{\star}$ represents the number of relativistic degrees of freedom, which depends on the details of particle physics and high-energy physics. In terms of finding out new physics, we can provide benchmark information about the reheating temperature for the various models of inflation studied in this paper, then we can further comment on aspects of the underlying microphysics associated with this equilibrium temperature for early universe physics.

    \item For the primordial power spectrum for the scalar modes, it is common to adopt the following parametrization:
    \begin{eqnarray}
        \Delta^2_{\zeta}(k)=\Delta^2_{\zeta}(k_{\rm cmb})\left(\frac{k}{k_{\rm cmb}}\right)^{n_s-1+\frac{\alpha_s}{2!}\ln\left(\frac{k}{k_{\rm cmb}}\right)+\frac{\beta_s}{3!}\ln^2\left(\frac{k}{k_{\rm cmb}}\right)+\cdots}
    \end{eqnarray}
where $\Delta^2_{\zeta}(k_{\rm cmb})$ represents the dimensionless scalar power spectrum, $n_s$ is the spectral tilt, $\alpha_s$ and $\beta_s$ are the running and running of the running of the spectral tilt, defined by the following expressions:
\begin{eqnarray}
    n_s&=&1+\left(\frac{d\ln \Delta^2_{\zeta}(k)}{d\ln k}\right)_{k=k_{\rm cmb}}=1+\left(\frac{d\ln \Delta^2_{\zeta}(k)}{dN}\right)_{N=N_{\rm cmb}},\\
    \alpha_s&=&\left(\frac{dn_s}{d\ln k}\right)_{k=k_{\rm cmb}}=\left(\frac{dn_s}{dN}\right)_{N=N_{\rm cmb}},\\
    \beta_s&=&\left(\frac{d^2 n_s}{d\ln k^2}\right)_{k=k_{\rm cmb}}=\left(\frac{d^2n_s}{dN^2}\right)_{N=N_{\rm cmb}}.
\end{eqnarray}
With the help of new ACT and DESI DR2 data, one can immediately constrain the value of the spectral tilt $n_s$ at the CMB scale. However, with our analysis, we try to provide estimates of the quantities $\alpha_s$ and $\beta_s$ for various models of inflation in the presence of non-minimal coupling, which we believe will provide significant information in terms of finding new physics. In this paper, we have explored this issue in detail.

    \item Last but not least, we comment on the field excursion $|\Delta\phi|=|\phi_{\rm cmb}-\phi_{\rm end}|$ which is described by the following modified Lyth bound as given by \cite{Choudhury:2024ezx,Choudhury:2023kam,Choudhury:2014kma,Choudhury:2015pqa},
    \begin{eqnarray}
        \frac{|\Delta\phi|}{M_{\rm pl}}&=&\sqrt{\frac{r}{8}}\frac{2}{\left(n_s-1-\frac{r}{8}\right)}\nonumber\\ &&\quad\quad\times\left|1-\exp\left(-|\Delta N|\frac{\left(n_s-1-\frac{r}{8}\right)}{2}\right)\right|,\quad
    \end{eqnarray}
    where at the end of inflation the field value is given by $\phi_{\rm end}$ corresponding to e-folds $N_{\rm end}$. When the exponential contribution is expanded in the Taylor series and we keep the first two terms, we can immediately get back the usual Lyth bound formula:
    \begin{eqnarray}
        \frac{|\Delta\phi|}{M_{\rm pl}}&=&\sqrt{\frac{r}{8}}|\Delta N|.
        \end{eqnarray}

    See Refs.~\cite{Lyth:1996im,Efstathiou:2005tq} for further discussion of the Lyth bound and its phenomenological implications for sub-Planckian and super-Planckian inflationary dynamics.
    
    Depending on the value of the field excursion, one can consider two possible branches, (1) where $|\Delta\phi|<M_{\rm pl}$, which is commonly known as sub-Planckian models of inflation where the Effective Field Theory (EFT) prescription is valid, and (2) where $|\Delta\phi|>M_{\rm pl}$, which is commonly known as super-Planckian models of inflation. Using our analysis performed in this paper we classify the field excursion values in the previously mentioned two regimes with the help of new ACT and DESI DR2 data. In terms of finding the outcomes of new physics, this information plays a significant role, which will be clearer as we proceed with our analysis.

    \item 
Primordial gravitational waves (PGWs), arising from quantum tensor fluctuations during inflation, offer a direct probe of the inflationary energy scale and gravitational interactions in the ultraviolet regime \cite{Guzzetti:2016mkm, Caprini:2018mtu}. In the standard single-field slow-roll inflation framework, the power spectrum of tensor modes is given by
\begin{equation}
\mathcal{P}_T(k) = \frac{2 H^2}{\pi^2 M_{\text{Pl}}^2},
\end{equation}
where $H$ is the Hubble parameter during inflation and $M_{\text{Pl}}$ is the reduced Planck mass. The tensor-to-scalar ratio is defined as
\begin{equation}
r \equiv \frac{\mathcal{P}_T(k_*)}{\mathcal{P}_\zeta(k_*)} \approx 16 \epsilon,
\end{equation}
with $\epsilon \equiv -\frac{\dot{H}}{H^2}$ being the first Hubble slow-roll parameter \cite{Liddle:2000cg}.

In non-minimally coupled models, where the inflaton $\phi$ interacts with gravity via a coupling term $\xi \phi^2 R$, the effective Planck mass becomes field-dependent:
\begin{equation}
M_{\text{eff}}^2(\phi) = M_{\text{Pl}}^2 + \xi \phi^2.
\end{equation}
This modifies the normalization of the tensor spectrum. After conformal transformation to the Einstein frame, the tensor amplitude transforms as
\begin{equation}
\mathcal{P}_T(k) = \frac{2 H^2}{\pi^2 M_{\text{eff}}^2(\phi)},
\end{equation}
leading to a generalized expression for $r$:
\begin{equation}
r = \frac{\mathcal{P}_T}{\mathcal{P}_\zeta} \propto \frac{H^2}{M_{\text{eff}}^2 \epsilon},
\end{equation}
introducing strong model dependence and breaking the canonical $r \simeq 16 \epsilon$ relation \cite{Kaiser:1994vs, Komatsu:1999mt}.

Furthermore, the tensor spectral index $n_T$ and its running $\alpha_T$ may deviate from scale invariance:
\begin{align}
n_T &\equiv \frac{d \ln \mathcal{P}_T(k)}{d \ln k} = -2\epsilon + \cdots, \\
\alpha_T &\equiv \frac{d n_T}{d \ln k},
\end{align}
with additional contributions depending on the form of $f(\phi)$ and higher-order slow-roll parameters \cite{Kallosh:2013hoa}.

Next-generation CMB experiments such as CMB-S4 and LiteBIRD aim to detect or constrain $r \lesssim 10^{-3}$ \cite{Abazajian:2016yjj, Hazumi:2019lys}, while space-based interferometers (LISA, DECIGO, BBO) will probe the tensor spectrum at higher frequencies \cite{Bartolo:2016ami}. These instruments can test both the inflationary scale and post-inflationary dynamics.

Gravitational waves can also be generated during the reheating phase. The frequency of the gravitational wave background sourced during reheating is approximately given by
\begin{equation}
f_{\text{reh}} \sim 10^7 \left( \frac{T_{\text{reh}}}{10^9\,\text{GeV}} \right) \text{Hz},
\end{equation}
which lies well above the CMB sensitivity range but within the reach of future detectors \cite{Easther:2006gt}. The spectrum is sensitive to the reheating temperature $T_{\text{reh}}$ and the details of inflaton decay, including potential resonance effects.

Moreover, non-minimal models with axion-like interactions $\phi F \tilde{F}$ may induce gauge field production that sources additional tensor perturbations:
\begin{equation}
\mathcal{P}_T^{\text{(gauge)}}(k) \propto e^{2\pi \xi},
\end{equation}
where $\xi \equiv \frac{\dot{\phi}}{2 f H}$ quantifies the strength of the coupling. These tensor modes can exhibit large amplitudes and a strongly blue-tilted spectrum, offering detection prospects at high frequencies \cite{Cook:2011hg, Barnaby:2011qe}.

In this work, we analyze current constraints on the tensor-to-scalar ratio $r$ using ACT and DESI DR2 data, and explore their implications for PGW observables in the context of non-minimally coupled inflation. This allows us to assess the viability of UV-complete models and post-inflationary dynamics consistent with high-precision cosmological data.

\end{enumerate}

\section{Numerical Analysis}
\label{sec:methodology}

In this section, we provide a description of our numerical pipeline, designed to compute inflationary observables for a wide array of single--field potentials with a non-minimal coupling to gravity \cite{Liddle:1994dx, Kaiser:1994vs, Faraoni:1996rf}.  Our framework accommodates various functional forms of the Jordan frame potential $V(\phi)$---including monomial, hilltop, plateau, and more exotic shapes---as well as different power-law coupling functions $f(\phi)=(\phi/M_{\rm pl})^n$.  Emphasis is placed on reproducibility, numerical stability, and flexibility, so that any new potential or coupling choice can be plugged into the same workflow.

\subsection{Jordan- to Einstein-Frame Transformation}
\label{ssec:frame_transform}

We begin with the Jordan frame Lagrangian,
\begin{equation}
  \mathcal{S}_J
  = \int d^4x\,\sqrt{-g}
    \Bigl[
      \tfrac{M_{\rm pl}^2}{2}\bigl(1+\xi\,f(\phi)\bigr)R
      - \tfrac12\,g^{\mu\nu}\partial_\mu\phi\,\partial_\nu\phi
      - V(\phi)
    \Bigr],
  \label{eq:action_jordan}
\end{equation}
where $M_{\rm pl}=2.43\times10^{18}\,$GeV is the reduced Planck mass, and $\xi$ parametrizes the strength of the non-minimal coupling to the Ricci scalar.  We typically choose $f(\phi)=(\phi/M_{\rm pl})^n$ for integer $n\ge1$, but the code accepts any smooth $f(\phi)$.

To recast the theory in a form amenable to standard slow-roll analysis, we perform the Weyl rescaling
\begin{equation}
  \hat g_{\mu\nu} = \Omega^2(\phi)\,g_{\mu\nu},
  \quad
  \Omega^2(\phi) \equiv 1 + \frac{\xi f(\phi)}{M_{\rm pl}^2},
  \label{eq:conformal}
\end{equation}
which renders the gravitational sector purely Einstein--Hilbert but introduces a non-trivial kinetic prefactor for $\phi$.  We restore canonical normalization by defining the new field variable
\begin{equation}
  \varphi(\phi)\;=\;\int_{\phi_{\rm ini}}^{\phi}
    \sqrt{\frac{1}{\Omega^2(\tilde\phi)}
          +\frac{3}{2}\Bigl(\frac{d\ln\Omega^2}{d\tilde\phi}\Bigr)^{\!2}
    }
    \;d\tilde\phi.
  \label{eq:varphi_def}
\end{equation}
Numerical evaluation of \eqref{eq:varphi_def} uses an adaptive Runge--Kutta--Fehlberg (RK45) integrator with relative tolerance $10^{-9}$ and absolute tolerance $10^{-12}$.  We discretize $\phi$ on a dense grid of $10^4$ points in $[\phi_{\rm end},\phi_{\rm ini}]$, where $\phi_{\rm ini}\sim30\,M_{\rm pl}$ is chosen to ensure the field begins well within the slow-roll regime.  If the integrator detects stiffness or fails to converge, step‐size control automatically refines the grid locally.

\subsection{Einstein-Frame Potential Construction}
\label{ssec:potential}

Under the same conformal transformation, the potential rescales as
\begin{equation}
  \hat V(\phi)
  = \frac{V(\phi)}{[\,\Omega^2(\phi)\,]^2}
  = \frac{V(\phi)}{\left[1+\dfrac{\xi f(\phi)}{M_{\rm pl}^2}\right]^2}.
  \label{eq:V_einstein}
\end{equation}
We then build $\hat V(\varphi)$ by inverting the numerical map $\varphi(\phi)$ via monotonic spline interpolation.  Concretely, we fit a cubic spline to the computed $(\phi,\varphi)$ data, evaluate its inverse on a regular $\varphi$ grid, and compute $\hat V(\varphi)$ by direct substitution.  This two‐step procedure maintains an overall accuracy better than $10^{-5}$ in $\hat V$.

\subsection{Calculation of Slow-Roll Parameters}
\label{ssec:slowroll}

Working with the canonical field $\varphi$, we compute derivatives of $\hat V(\varphi)$ using a five-point central finite-difference stencil.  Denoting $d/d\varphi$ by primes, the Hubble slow-roll parameters are
\begin{align}
  \epsilon(\varphi)&=\tfrac12\Bigl(\tfrac{\hat V'}{\hat V}\Bigr)^2,
  &\eta(\varphi)&=\tfrac{\hat V''}{\hat V},
  \notag\\
  \lambda_3(\varphi)&=\tfrac{\hat V'\,\hat V'''}{\hat V^2},
  &\lambda_4(\varphi)&=\tfrac{\hat V''\,\hat V''''}{\hat V^2}.
  \label{eq:slowroll_params}
\end{align}
Here $\lambda_3$ denotes the third potential slow-roll combination. We use this symbol, rather than $\xi^2$, to avoid any ambiguity with the non-minimal gravitational coupling $\xi$.
From these we derive the primary spectral observables:
\begin{align}
  n_s(\varphi)&=1 - 6\,\epsilon + 2\,\eta,
  \label{eq:ns}\\
  \alpha_s(\varphi)&=16\,\epsilon\,\eta - 24\,\epsilon^2 - 2\,\lambda_3,
  \label{eq:alpha_s}\\
  \beta_s(\varphi)&=-192\,\epsilon^3 +192\,\epsilon^2\eta -32\,\epsilon\,\eta^2
     -24\,\epsilon\,\lambda_3 +2\,\eta\,\lambda_3 +2\,\lambda_4.
  \label{eq:beta_s}
\end{align}
To monitor numerical error, we periodically compare to a three-point stencil; any point exceeding a relative discrepancy of $10^{-3}$ triggers a local refinement of the $\varphi$ grid.

\subsection{Termination of Inflation and \emph{e}-fold Integration}
\label{ssec:efolds}

Inflation ends at $\varphi_{\rm end}$ defined by $\epsilon(\varphi_{\rm end})=1$.  We bracket this root by scanning for the first grid point satisfying $\epsilon\ge1$ and then apply Brent's method to locate $\varphi_{\rm end}$ to within $10^{-4}\,M_{\rm pl}$.  The cumulative number of \emph{e}-folds from $\varphi$ to $\varphi_{\rm end}$ is
\begin{equation}
  N(\varphi)
  =\frac{1}{M_{\rm pl}}
   \int_{\varphi_{\rm end}}^{\varphi}
     \frac{d\tilde\varphi}{\sqrt{2\,\epsilon(\tilde\varphi)}},
  \label{eq:efolds}
\end{equation}
evaluated by an adaptive composite trapezoidal rule with tolerance $10^{-6}$ in $N$ to ensure precise inversion when solving $N(\varphi_\star)=N_\star$.

\subsection{Extraction of Observables}
\label{ssec:observables}

For pivot \emph{e}-fold values $N_\star=50$ and $60$, we invert Eq.~\eqref{eq:efolds} to find the corresponding field values $\varphi_\star$.  At each pivot we then evaluate:
\[
  n_s = n_s(\varphi_\star),\quad
  r = 16\,\epsilon(\varphi_\star),\quad
  \alpha_s = \alpha_s(\varphi_\star),\quad
  \beta_s = \beta_s(\varphi_\star).
\]
The energy scale of inflation is inferred via:
\begin{equation}
V_{\rm inf}^{1/4} = \left( \frac{3\pi^2 r A_s}{2} \right)^{1/4} M_{\rm pl},
\label{eq:Vinf}
\end{equation}
with $A_s \simeq 2.1 \times 10^{-9}$ from Planck \cite{Planck:2018jri}. The reheating temperature is estimated as
\begin{equation}
  T_{\rm reh}=\Bigl(\tfrac{90}{\pi^2g_*}\Bigr)^{\!1/4}\sqrt{H_{\rm inf}M_{\rm pl}},
  \label{eq:Treh}
\end{equation}
where $g_*=106.75$ is the effective count of relativistic degrees of freedom. 

Equation~\eqref{eq:Treh} is used in this work as an instantaneous-reheating benchmark estimate. The quoted values of $T_{\rm reh}$ should therefore be interpreted as indicative reference scales rather than unique reheating predictions. In a more general reheating scenario with finite duration and effective equation-of-state parameter $w_{\rm reh}$, both the reheating temperature and the mapping between horizon exit and the number of e-folds $N_\star$ would be modified. Consequently, the high-frequency tail of the primordial gravitational-wave spectrum can also shift, especially for modes re-entering the horizon during reheating.

Uncertainties in $n_s$ and $r$ due to numerical tolerances are estimated by propagating a $\pm5\%$ variation in integration tolerances through to the final values.

\subsection{Field Excursion and Modified Lyth Bound}
\label{ssec:lyth}

The total field excursion in canonical coordinates is
\[
  \Delta\varphi=\bigl|\varphi(\phi_\star)-\varphi_{\rm end}\bigr|.
\]
We compare this to the generalized Lyth bound \cite{Lyth:1996im, Efstathiou:2005tq}:
\begin{equation}
  \frac{\Delta\varphi}{M_{\rm pl}}
  =\sqrt{\frac{r}{8}}\,\frac{2}{|n_s-1-r/8|}
   \Bigl[1-\exp\bigl(-|\Delta N|\tfrac{|n_s-1-r/8|}{2}\bigr)\Bigr],
  \label{eq:modified_lyth}
\end{equation}
and verify that in the limit $|n_s-1-r/8|\ll1$ it reduces to the conventional $\sqrt{r/8}\,\Delta N$ within $1\%$ error.

\subsection{Inverse Coupling Scan: Numerical Reconstruction of $\xi$}
\label{ssec:inverse_scan}

Having obtained the inflationary observables $n_s(\xi)$ and $r(\xi)$ as numerically evaluated functions of the non-minimal coupling parameter $\xi$, we now address the inverse problem: for a given target pair $(n_s^{\rm tgt}, r^{\rm tgt})$ motivated by observational constraints, determine the corresponding $\xi^\star$ such that the model reproduces these values within a specified accuracy.

To formulate this inversion quantitatively, we define a chi-squared loss functional over the observable space:
\begin{equation}
  \chi^2(\xi) \equiv 
  \left[n_s(\xi) - n_s^{\rm tgt}\right]^2 + 
  \left[r(\xi) - r^{\rm tgt}\right]^2,
  \label{eq:chi2}
\end{equation}
which acts as a scalar objective function to be minimized. The functional $\chi^2(\xi)$ is non-negative and achieves its global minimum $\chi^2_{\min} = 0$ if and only if both observables are matched exactly. In practice, this rarely occurs due to numerical errors and model degeneracies, so we accept approximate matches with $\chi^2_{\min} \lesssim 10^{-6}$.

The search domain for $\xi$ is initially bounded within a conservative physically-motivated interval $\xi \in [10^{-3}, 10]$, wide enough to encapsulate weak and strong coupling regimes, yet avoiding numerical instabilities from extreme curvature of the field redefinition in the Einstein frame \cite{Kaiser:1994vs, Bezrukov:2007ep}.

Minimization of Eq.~\eqref{eq:chi2} is carried out using Brent's method \cite{Brent:1973algo}, a robust root-bracketing scalar minimization algorithm that requires only function evaluations and no derivatives. The implementation is via the \texttt{scipy.optimize.minimize\_scalar} routine with absolute tolerance $\epsilon = 10^{-4}$ to ensure convergence precision in the retrieved $\xi^\star$. We record the solution only if:
\begin{enumerate}
    \item The optimizer converges to an interior point $\xi^\star \in (10^{-3}, 10)$,
    \item The minimum $\chi^2(\xi^\star) < 10^{-6}$,
    \item The final solution satisfies consistency checks on $\hat{V}(\varphi)$ monotonicity and slow-roll validity.
\end{enumerate}

To systematically characterize the space of viable models, we define a two-dimensional rectangular grid in the $(n_s^{\rm tgt}, r^{\rm tgt})$ plane:
\begin{align}
  n_s^{\rm tgt} &\in [0.967, 0.975] \quad \text{with} \quad \Delta n_s = 10^{-3}, \\
  r^{\rm tgt}   &\in [0.001, 0.030] \quad \text{with} \quad \Delta r = 10^{-3},
\end{align}
covering 81 points in total. These intervals are chosen based on the most recent cosmological bounds: the joint Planck+ACT (P-ACT) likelihood gives $n_s = 0.9709 \pm 0.0038$ at 68\% CL \cite{ACT:2025fju}, and the inclusion of DESI BAO and CMB lensing (P-ACT-LB) yields $n_s = 0.9743 \pm 0.0034$, marking a $\sim2\sigma$ deviation from Planck-only results \cite{ACT:2025tim}.

For each grid point, the optimizer attempts to find $\xi^\star$ minimizing $\chi^2$. If the returned minimum lies on the boundary of the interval, i.e., $\xi^\star = 10^{-3}$ or $10$, the domain is dynamically extended to $\xi \in [10^{-5}, 10^2]$ and the search is repeated. If still unresolved, the scan returns $\xi^\star = \mathrm{None}$, indicating model incompatibility or strong-coupling breakdown. 

Each successful minimization produces a tuple $(n_s^{\rm tgt}, r^{\rm tgt}, \xi^\star)$ stored in a structured database. We partition the resulting dataset into:
\begin{itemize}
  \item $\mathcal{S}_{\rm valid}$: configurations with $\xi^\star \in (10^{-5}, 10^2)$ and $\chi^2 < 10^{-6}$;
  \item $\mathcal{S}_{\rm partial}$: fits with $10^{-6} \leq \chi^2 < 10^{-3}$;
  \item $\mathcal{S}_{\rm invalid}$: boundary or divergence cases.
\end{itemize}

Post-processing filters out pathological cases, such as discontinuities in $\varphi(\phi)$ due to numerical stiffness or abrupt non-monotonicity in $\hat{V}(\varphi)$. Diagnostic logging captures all failures for reproducibility and further refinement.

This inversion scan constitutes a robust numerical map from observable constraints to theoretically permissible coupling values $\xi$, shedding light on the viability of various inflationary potentials in the context of current and future precision cosmological data.

\section{Non-minimally Coupled Single-Field Inflation Models}
\label{sec:models}

In this section, we present the theoretical formulation underlying our analysis of single-field inflationary models with a non-minimal coupling to gravity. The models are defined in the Jordan frame by the following action:

\begin{equation}
S = \int d^4x \sqrt{-g} \left[ \frac{M_{\rm pl}^2}{2} \left( 1 + \xi\, f(\phi) \right) R - \frac{1}{2} g^{\mu\nu} \partial_\mu \phi \, \partial_\nu \phi - V(\phi) \right],
\label{eq:jordan_action}
\end{equation}

where $\phi$ is the scalar inflaton field, $R$ is the Ricci scalar derived from the Jordan frame metric $g_{\mu\nu}$, $M_{\rm pl} = (8\pi G)^{-1/2}$ is the reduced Planck mass, and $\xi$ is a dimensionless parameter quantifying the non-minimal coupling between the scalar field and the curvature scalar. The function $f(\phi)$ encodes model-dependent features, with $f(\phi) = \phi^2$ often taken as a benchmark \cite{Fakir:1990eg, Bezrukov:2007ep}. The action reduces to the canonical minimally coupled form in the limit $\xi \to 0$.

To analyze the inflationary dynamics in a frame with canonical gravitational and kinetic terms, a Weyl (conformal) transformation is performed on the metric:

\begin{equation}
g_{\mu\nu} \rightarrow \hat{g}_{\mu\nu} = \Omega^2(\phi) g_{\mu\nu}, \qquad \Omega^2(\phi) \equiv 1 + \frac{\xi f(\phi)}{M_{\rm pl}^2}.
\label{eq:conformal_transformation}
\end{equation}

This transformation maps the theory to the Einstein frame, in which the gravitational sector assumes the canonical Einstein--Hilbert form, but at the expense of introducing noncanonical kinetic terms for the scalar field \cite{Kaiser:1995nv}.

In the Einstein frame, the kinetic term of the scalar field becomes nontrivial, and a field redefinition is required to recover canonical normalization. The canonically normalized field $\varphi$ is defined through the field-space metric:

\begin{equation}
\left( \frac{d\varphi}{d\phi} \right)^2 = \frac{1}{\Omega^2(\phi)} \left[ 1 + \frac{3 M_{\rm pl}^2}{2} \left( \frac{d \ln \Omega^2(\phi)}{d\phi} \right)^2 \right],
\label{eq:field_redefinition}
\end{equation}

leading to a nontrivial kinetic mixing arising from the curvature of the scalar-field manifold induced by the conformal transformation \cite{Kallosh:2013hoa, Galante:2014ifa}. The scalar potential transforms according to

\begin{equation}
\hat{V}(\varphi) = \frac{V(\phi(\varphi))}{\Omega^4(\phi(\varphi))},
\label{eq:einstein_potential}
\end{equation}

resulting in an effective potential in the Einstein frame that governs the inflationary dynamics of the canonically normalized field $\varphi$.

Thus, the analysis of inflationary observables across a range of non-minimally coupled models is unified within a single framework, with model-dependence entirely encapsulated in the Jordan frame functions $V(\phi)$ and $f(\phi)$.

\vspace{0.5em}
For each model under consideration, we numerically compute the inflationary observables for a broad logarithmic span of the non-minimal coupling $\xi \in [10^{-3}, 10^2]$. Calculations are performed for benchmark values of the number of e-folds before the end of inflation, $N_\star = 50$ and $60$, reflecting plausible reheating histories \cite{Liddle:2003as}.

We evaluate the following key inflationary observables:

\begin{itemize}
\item Scalar spectral index $n_s$,
\item Tensor-to-scalar ratio $r$,
\item Running of the scalar spectral index $\alpha_s \equiv \frac{d n_s}{d \ln k}$,
\item Running of the running $\beta_s \equiv \frac{d^2 n_s}{d (\ln k)^2}$,
\item Einstein frame field excursion $\Delta \varphi / M_{\rm pl}$,
\item Hubble scale during inflation $H_{\rm inf}$,
\item Reheating temperature $T_{\rm reh}$.
\end{itemize}

All computations are performed using a numerical slow-roll approximation pipeline \cite{Martin:2013tda}, with the observables expressed in terms of the Einstein frame slow-roll parameters:

\begin{align}
\epsilon(\varphi) &\equiv \frac{M_{\rm pl}^2}{2} \left( \frac{\hat{V}'(\varphi)}{\hat{V}(\varphi)} \right)^2, \label{eq:epsilon} \\
\eta(\varphi) &\equiv M_{\rm pl}^2 \frac{\hat{V}''(\varphi)}{\hat{V}(\varphi)}, \label{eq:eta} \\
\lambda_3(\varphi) &\equiv M_{\rm pl}^4 \frac{\hat{V}'(\varphi)\hat{V}'''(\varphi)}{\hat{V}^2(\varphi)}, \label{eq:lambda3_slowroll}
\end{align}

where primes denote derivatives with respect to the canonical field $\varphi$. The notation $\lambda_3$ is used for the third potential slow-roll combination in order to avoid confusion with the non-minimal coupling parameter $\xi$.

The spectral observables are then computed to second order in slow roll:

\begin{align}
n_s &= 1 - 6\epsilon + 2\eta, \label{eq:ns_model} \\
r &= 16\epsilon, \label{eq:r} \\
\alpha_s &= 16\epsilon\eta - 24\epsilon^2 - 2\lambda_3, \label{eq:alphas}
\end{align}

while the running of the running $\beta_s$ is derived by differentiation of $\alpha_s$ with respect to $\ln k$.

The Hubble scale during inflation is obtained from the Einstein frame Friedmann equation at horizon crossing:

\begin{equation}
H_{\rm inf} = \sqrt{\frac{\hat{V}(\varphi_\star)}{3 M_{\rm pl}^2}},
\label{eq:hubble}
\end{equation}

where $\varphi_\star$ is the field value at $N_\star$ e-folds before the end of inflation.

Assuming perturbative decay of the inflaton, the reheating temperature is given by

\begin{equation}
T_{\rm reh} \simeq \left( \frac{90}{\pi^2 g_*} \right)^{1/4} \sqrt{\Gamma_\phi M_{\rm pl}},
\label{eq:reheat_temp}
\end{equation}

where $\Gamma_\phi$ is the total inflaton decay width and $g_*$ denotes the effective number of relativistic degrees of freedom at reheating \cite{KolbTurner:1990}.

Equation~\eqref{eq:reheat_temp} represents the more general perturbative-decay estimate for reheating. In the numerical analysis presented in this work, however, we do not attempt a model-dependent determination of $\Gamma_\phi$. Instead, for the purpose of a uniform model-by-model comparison, all quoted values of $T_{\rm reh}$ are evaluated in the instantaneous-reheating limit and should be interpreted as benchmark reference scales.

This formalism enables rigorous, quantitative comparison of a wide class of non-minimally coupled inflationary models in terms of their inflationary and post-inflationary predictions.

\begin{table*}
\begin{center}
\resizebox{\textwidth}{!}{%
\begin{tabular}{lccccccccc}
\noalign{\hrule\vskip 2pt} \noalign{\hrule\vskip 3pt}
\textbf{Inflationary Model} & \boldmath$f(\phi) = (\phi/M_{\rm Pl})^k$ & \boldmath$\xi$ & \boldmath$r$ & \boldmath$n_s$ & \boldmath$\alpha_s$ & \boldmath$\beta_s$ & \boldmath$H_{\mathrm{inf}}$ [$10^{16}$ GeV] & \boldmath$T_{\mathrm{reh}}$ [$10^{16}$ GeV] & \boldmath$\frac{|\Delta\phi|}{M_{\rm Pl}}$ \\
\noalign{\hrule\vskip 3pt}
Starobinsky              & $k = 2$ & $0.01$ -- $10$       & 0.00299 & 0.96767 & $-5.29\times10^{-4}$ & $2.95\times10^{-4}$ & 0.144 & 0.320 & 0.3594 \\
Higgs Inflation          & $k = 2$ & $0.01$ -- $10$       & 0.00349 & 0.96756 & $-5.32\times10^{-4}$ & $2.48\times10^{-4}$ & 0.155 & 0.333 & 0.3870 \\
T-model ($m = 1$)        & $k = 1$ & $0.026$ -- $100$     & 0.00055 & 0.96669 & $-5.55\times10^{-4}$ & $2.13\times10^{-3}$ & 0.0618 & 0.210 & 0.1523 \\
Hilltop (Quartic)        & $k = 1$ & $0.222$ -- $9.983$   & 0.00926 & 0.97352 & $-4.51\times10^{-4}$ & $5.32\times10^{-5}$ & 0.253 & 0.424 & 0.7046 \\
D-brane Inflation ($p=2$)& $k = 1$ & $1.502$ -- $1.997$   & 0.00101 & 0.96309 & $-5.91\times10^{-4}$ & $1.29\times10^{-3}$ & 0.0836 & 0.244 & 0.1913 \\
\noalign{\vskip 2pt\hrule}
\end{tabular}%
}
\end{center}
\caption{\label{tab:inflation_models_summary} 
Inflationary models considered in this work for $N = 60$. The non-minimal coupling function is written as $f(\phi)=(\phi/M_{\rm Pl})^k$. 
The predictions in the $(n_s, r)$ plane are shown in Figures~\ref{fig:r_ns2} and~\ref{fig:r_ns1}: 
Figure~\ref{fig:r_ns2} shows the trajectories of the Starobinsky and Higgs models, while 
Figure~\ref{fig:r_ns1} shows those of the T-model, Hilltop, and D-brane models across their respective $\xi$ ranges.
The quoted reheating temperatures are benchmark values obtained in the instantaneous-reheating limit. For a finite reheating duration with effective equation-of-state parameter $w_{\rm reh}$, both $T_{\rm reh}$ and the high-frequency part of the primordial GW spectrum may shift; therefore these values should be interpreted as reference scales for model comparison rather than unique reheating predictions.}
\end{table*}
\begin{table*}
\centering
\begin{tabular}{lccccccc}
\hline
\textbf{Model} & $\boldsymbol{\xi}$ & $\boldsymbol{n_s}$ & $\boldsymbol{r}$ & $\boldsymbol{\Delta \phi / M_{\mathrm{Pl}}}$ & $\boldsymbol{H_{\mathrm{inf}}\,(\mathrm{GeV})}$ & $\boldsymbol{T_{\mathrm{reh}}\,(\mathrm{GeV})}$ & $\boldsymbol{\Omega_{\mathrm{GW}} h^2}$ \\
\hline
\textbf{Starobinsky} & 0.01 & 0.9628 & $4.10 \times 10^{-3}$ & 0.39  & $1.8 \times 10^{14}$ & $2.0 \times 10^{14}$ & --- \\
                     & 1    & 0.9630 & $3.90 \times 10^{-3}$ & 0.35  & $1.8 \times 10^{14}$ & $2.0 \times 10^{14}$ & $1.716 \times 10^{-13}$ \\
                     & 10   & 0.9631 & $3.85 \times 10^{-3}$ & a0.33  & $1.8 \times 10^{14}$ & $2.0 \times 10^{14}$ & --- \\
\hline
\textbf{Higgs}       & 0.01 & 0.9566 & $6.39 \times 10^{-2}$ & 1.1354 & $6.65 \times 10^{13}$ & $6.88 \times 10^{15}$ & --- \\
                     & 0.10 & 0.9602 & $1.54 \times 10^{-2}$ & 0.5280 & $3.26 \times 10^{13}$ & $4.82 \times 10^{15}$ & --- \\
                     & 1    & 0.9676 & $3.49 \times 10^{-3}$ & 0.3870 & $1.55 \times 10^{13}$ & $3.33 \times 10^{15}$ & $1.460 \times 10^{-13}$ \\
                     & 10   & 0.9605 & $4.49 \times 10^{-3}$ & 0.3617 & $1.76 \times 10^{13}$ & $3.54 \times 10^{15}$ & --- \\
\hline
\textbf{T-Model} ($\alpha$-attractor) & 16.67 & 0.9421 & $9.7 \times 10^{-2}$ & 1.10  & $7.64 \times 10^{15}$ & $8.20 \times 10^{13}$ & --- \\
                                     & 1.0   & 0.9667 & $2.96 \times 10^{-3}$ & 0.35  & $3.19 \times 10^{15}$ & $1.43 \times 10^{13}$ & $1.239 \times 10^{-13}$ \\
                                     & 0.0017 & 0.9666 & $6.0 \times 10^{-6}$ & 0.015 & $7.48 \times 10^{14}$ & $6.21 \times 10^{11}$ & --- \\
\hline
\textbf{Hilltop Quartic} & 6.9932 & 0.96931 & $3.26 \times 10^{-3}$ & 0.3877 & $1.50 \times 10^{13}$ & $3.27 \times 10^{15}$ & --- \\
                         & 3.3620 & 0.97142 & $4.18 \times 10^{-3}$ & 0.4587 & $1.70 \times 10^{13}$ & $3.48 \times 10^{15}$ & $1.749 \times 10^{-13}$ \\
                         & 9.9488 & 0.96858 & $3.08 \times 10^{-3}$ & 0.3714 & $1.46 \times 10^{13}$ & $3.22 \times 10^{15}$ & --- \\
\hline
\textbf{D-brane Inflation} & 1.986  & 0.96305 & $1.014 \times 10^{-3}$ & 0.1915 & $8.37 \times 10^{12}$ & $2.44 \times 10^{15}$ & --- \\
                           & 1.997  & 0.96309 & $1.011 \times 10^{-3}$ & 0.1913 & $8.36 \times 10^{12}$ & $2.44 \times 10^{15}$ & $4.231 \times 10^{-14}$ \\
                           & 1.9974 & 0.96306 & $1.012 \times 10^{-3}$ & 0.1914 & $8.37 \times 10^{12}$ & $2.44 \times 10^{15}$ & --- \\
\hline
\end{tabular}
\caption{Summary of primordial gravitational wave signals, quantified by the tensor-to-scalar ratio $r$, and reheating temperatures $T_{\mathrm{reh}}$ across various inflationary models for selected values of the non-minimal coupling parameter $\xi$. The table also lists the scalar spectral index $n_s$, the inflaton field excursion $\Delta \phi$ normalized by the reduced Planck mass $M_{\mathrm{Pl}}$, the inflationary Hubble scale $H_{\mathrm{inf}}$ (in GeV), and the dimensionless energy density of primordial gravitational waves $\Omega_{\mathrm{GW}} h^2$ evaluated at a characteristic frequency $f^*$. All observables are evaluated at $N_e = 60$ e-folds before the end of inflation. All values of $T_{\rm reh}$ quoted in this table are obtained under the assumption of instantaneous reheating and should therefore be interpreted as benchmark estimates. A finite reheating phase with general $w_{\rm reh}$ can modify both $T_{\rm reh}$ and the high-frequency primordial GW spectrum.}
\label{tab:GW_Treh_summary}
\end{table*}

\begin{table*}
\centering
\renewcommand{\arraystretch}{1.4}
\begin{tabular}{lcc|cc}
\hline
\textbf{Model} 
& $n_s$ & $\alpha_s$ 
& $n_s$ & $\alpha_s$ \\
& ($N = 50$) & ($N = 50$) 
& ($N = 60$) & ($N = 60$) \\
\hline
Starobinsky         & $0.961345$ & $-7.57 \times 10^{-4}$ & $0.967668$ & $-5.29 \times 10^{-4}$ \\
Higgs               & $0.961187$ & $-7.63 \times 10^{-4}$ & $0.967563$ & $-5.32 \times 10^{-4}$ \\
D-Brane             & $0.955997$ & $-8.65 \times 10^{-4}$ & $0.963085$ & $-5.91 \times 10^{-4}$ \\
Hilltop (Quartic)   & $0.968125$ & $-6.51 \times 10^{-4}$ & $0.973524$ & $-4.51 \times 10^{-4}$ \\
$\alpha$-attractor  & $0.960035$ & $-7.99 \times 10^{-4}$ & $0.966694$ & $-5.55 \times 10^{-4}$ \\
\hline
\end{tabular}
\caption{Values of the scalar spectral index $n_s$ and the running of the spectral index $\alpha_s$ for various inflationary models evaluated at $N = 50$ and $N = 60$ e-folds.}
\label{tab:ns_alpha_values}
\end{table*}

\begin{table}[h!]
\centering
\begin{tabular}{ccccc}
\hline
$\xi$ & $n_s$ & $r$ & $\alpha_s$ & $\Delta \phi / M_{\mathrm{Pl}}$ \\
\hline
0.01 & 0.9628 & $4.10 \times 10^{-3}$ & $-6.00 \times 10^{-4}$ & 0.390 \\
1 & 0.9630 & $3.90 \times 10^{-3}$ & $-5.90 \times 10^{-4}$ & 0.350 \\
10 & 0.9631 & $3.85 \times 10^{-3}$ & $-5.80 \times 10^{-4}$ & 0.330 \\
\hline
\end{tabular}
\caption{Inflationary observables evaluated at $N_e = 50$ e-folds before the end of inflation for representative values of the non-minimal coupling $\xi$. Running of the running $\beta_s \sim (2.8\,\text{–}\,3.0) \times 10^{-5}$ is not shown in the table but included in the numerical analysis.}
\label{tab:results_xi}
\end{table}

\section{Models under Study}
\label{sec:models_under_study}

\begin{figure*}[htbp]
    \centering
    \includegraphics[width=\textwidth, height=0.4\textwidth]{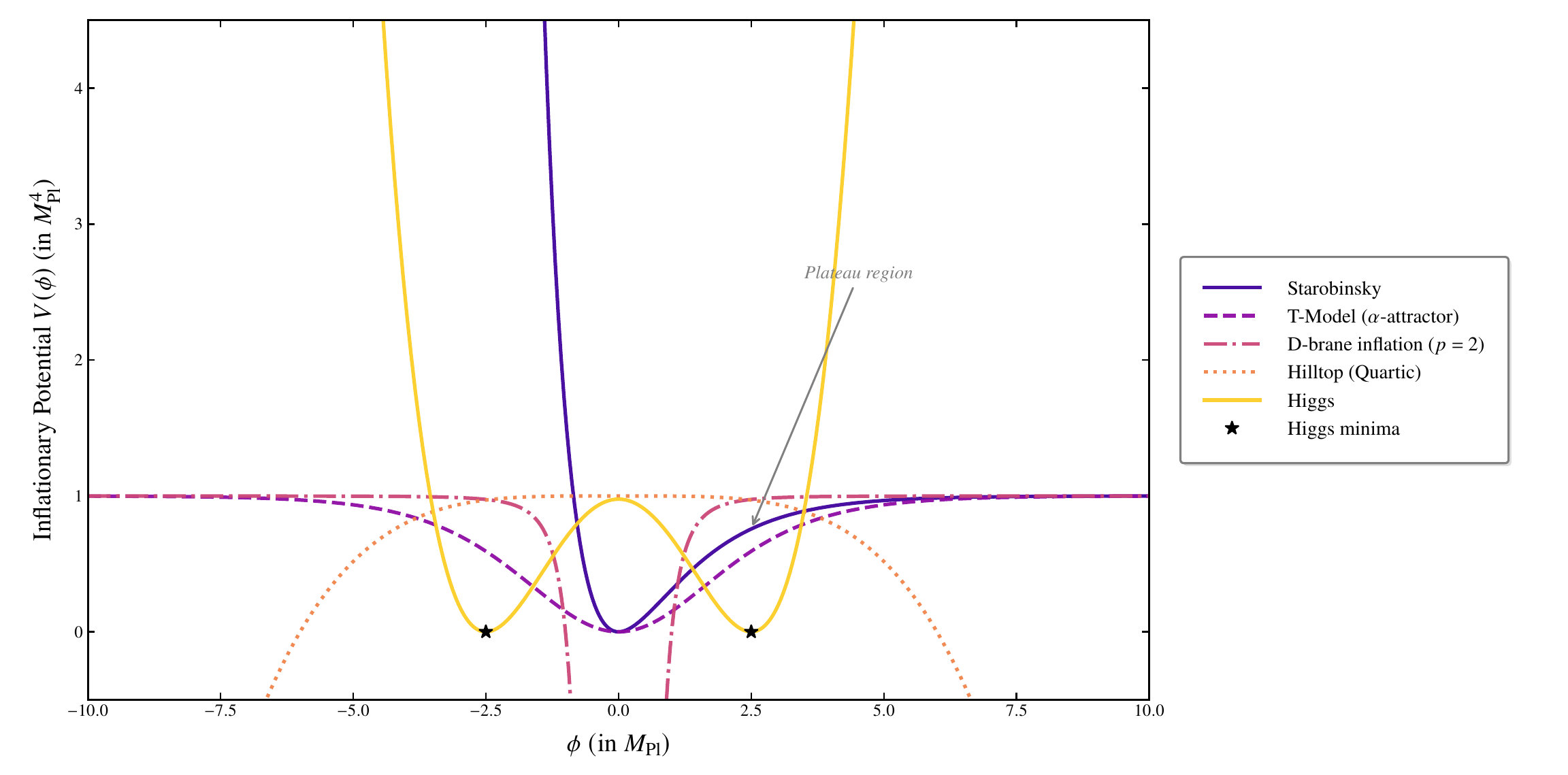}
    \caption{Comparison of inflationary potentials $V(\phi)$ for the models considered in this work. The plateau structure in the Starobinsky and $\alpha$-attractor T-Model potentials facilitates slow-roll inflation with suppressed tensor-to-scalar ratio $r$. The Higgs potential exhibits spontaneous symmetry breaking with characteristic minima. The inset highlights representative features such as plateaus and minima over the inflaton field range, expressed in Planck units.}
    \label{fig:potentials}
\end{figure*}

In the following subsections we use the unified Jordan-to-Einstein frame machinery of Sections~\ref{sec:methodology}--\ref{sec:models}. To avoid unnecessary duplication, the model descriptions emphasize only the potential-specific ingredients and phenomenological consequences; all slow-roll observables are evaluated through the common numerical pipeline described above. Although the general conformal transformation and slow-roll pipeline have already been described in Sections~\ref{sec:methodology}--\ref{sec:models}, we retain the leading analytic relations for each benchmark model in order to make clear which model-specific potential, limiting regime, and parameter dependence enter the numerical scan. The formulae in the following subsections should therefore be understood as compact model-specific specializations of the common framework rather than as independent computational procedures.

We investigate five single-field inflationary models, each chosen for their strong theoretical motivation and distinctive potential landscape. These include the Starobinsky model \cite{Starobinsky:1980te}, the $\alpha$-attractor T-Model \cite{Kallosh:2013hoa, Kallosh:2013lkr}, D-brane inflation with power-law exponent $p=2$ \cite{Dvali:1998pa, Burgess:2001fx}, the quartic Hilltop model \cite{Boubekeur:2005zm}, and a Higgs-like potential featuring spontaneous symmetry breaking \cite{Bezrukov:2007ep}.

The \textbf{Starobinsky model} originates from an $f(R)$ modification to the Einstein-Hilbert action, specifically $f(R) = R + R^2 / (6 M^2)$, where $M$ is a mass scale. This model is dynamically equivalent to a scalar-tensor theory with a potential that asymptotically approaches a plateau at large field values in the Einstein frame, providing robust predictions for $n_s$ and a characteristically small $r$ \cite{Starobinsky:1980te}.

The \textbf{$\alpha$-attractor T-Model} arises from superconformal supergravity models in which the kinetic term exhibits a pole structure. The Einstein frame potential takes the form $V(\phi) = V_0 \tanh^2(\phi / \sqrt{6\alpha})$, where $\alpha$ controls the width of the plateau. For small $\alpha$, the predictions approach universal attractor values regardless of the detailed microphysics \cite{Kallosh:2013hoa, Kallosh:2013lkr}.

The \textbf{D-brane inflation model} with $p = 2$ is motivated by string theory scenarios in which a D3-brane moves in a warped throat geometry. The resulting potential typically takes the form $V(\phi) = V_0 \left(1 - (\mu/\phi)^p \right)$, with $\mu$ controlling the scale of brane separation. For $p = 2$, this yields a steeper potential that can support inflation under specific initial conditions \cite{Dvali:1998pa, Burgess:2001fx}.

The \textbf{quartic Hilltop model} is a small-field inflation scenario characterized by a potential of the form $V(\phi) = V_0 \left(1 - (\phi/\mu)^4 \right)$. Inflation occurs near the local maximum at $\phi = 0$, where the potential is sufficiently flat for small values of $\phi$. This setup can generate red-tilted spectra with relatively small field excursions, depending on the parameter $\mu$ \cite{Boubekeur:2005zm}.

Finally, the \textbf{Higgs-like potential} we consider is given by $V(\phi) = \lambda (\phi^2 - v_h^2)^2 / 4$, which exhibits spontaneous symmetry breaking with degenerate minima at $\phi = \pm v_h$. When augmented by a non-minimal coupling to gravity, this potential becomes viable for inflation, particularly in the high-field regime where the effective potential flattens out \cite{Bezrukov:2007ep}.

Figure~\ref{fig:potentials} displays the shape of each potential over a representative range of inflaton field values, expressed in reduced Planck units ($M_{\rm pl} = 1$). Notably, the Starobinsky and $\alpha$-attractor models exhibit asymptotic plateaus, leading to slow-roll behavior with strongly suppressed tensor amplitudes. In contrast, the D-brane and Hilltop models display steep or unstable regions near the origin. The Higgs-like potential shows characteristic double-well symmetry, with inflation driven at large field values far from the vacuum expectation value.

These structural differences directly impact inflationary observables such as the scalar spectral index $n_s$, the tensor-to-scalar ratio $r$, and the total field excursion $\Delta\varphi$, all of which are analyzed in the following sections through both analytical and numerical techniques.

\subsection{The Starobinsky Model}
\label{ssec:starobinsky}

The Starobinsky model \cite{Starobinsky:1980te} represents a seminal inflationary scenario driven by higher-derivative corrections to the Einstein-Hilbert action. It originates from the addition of a quadratic Ricci scalar term to the gravitational Lagrangian, resulting in the action
\begin{equation}
    S = \frac{M_{\rm pl}^2}{2} \int d^4x \, \sqrt{-g} \left[ R + \frac{1}{6 M^2} R^2 \right],
    \label{eq:starobinsky_action}
\end{equation}
where $M_{\rm pl} \equiv (8\pi G)^{-1/2}$ is the reduced Planck mass, $R$ is the Ricci scalar, and $M$ is a mass parameter controlling the scale of the $R^2$ correction. The $R^2$ term arises naturally in the one-loop effective action of quantum-corrected gravity and provides a natural mechanism for driving cosmic inflation \cite{Starobinsky:1980te, Sebastiani_2014}.

The action \eqref{eq:starobinsky_action} leads to fourth-order equations of motion, but can be recast into a second-order scalar-tensor theory via the method of Lagrange multipliers. Introducing an auxiliary field $\chi$, the action becomes
\begin{equation}
    S = \frac{M_{\rm pl}^2}{2} \int d^4x \sqrt{-g} \left[ \left(1 + \frac{\chi}{3 M^2} \right) R - \frac{\chi^2}{6 M^2} \right],
    \label{eq:auxiliary_action}
\end{equation}
which resembles a Brans-Dicke-type theory with vanishing Brans-Dicke parameter $\omega = 0$.

To obtain a canonical scalar field description, we perform a Weyl (conformal) transformation
\begin{equation}
    g_{\mu\nu} \rightarrow \hat{g}_{\mu\nu} = \Omega^2(\phi) g_{\mu\nu}, \quad \Omega^2(\phi) = e^{\sqrt{\frac{2}{3}} \frac{\phi}{M_{\rm pl}}},
\end{equation}
and identify the canonical field $\phi$ via
\begin{equation}
    \phi = \sqrt{\frac{3}{2}} M_{\rm pl} \ln \left(1 + \frac{\chi}{3 M^2} \right).
\end{equation}
This procedure maps the original $f(R)$ theory to an Einstein frame scalar field model with canonical kinetic term and potential
\begin{equation}
    V(\phi) = \frac{3}{4} M_{\rm pl}^2 M^2 \left(1 - e^{-\sqrt{\frac{2}{3}} \frac{\phi}{M_{\rm pl}}} \right)^2,
    \label{eq:starobinsky_potential}
\end{equation}
which features a plateau at large $\phi$ and supports slow-roll inflation. This functional form ensures an exponentially flat potential, allowing the universe to undergo prolonged inflation with a small tensor-to-scalar ratio $r$ and nearly scale-invariant spectral index $n_s$.

\medskip

The slow-roll parameters,
\begin{align}
    \epsilon(\phi) &= \frac{M_{\rm pl}^2}{2} \left( \frac{V'(\phi)}{V(\phi)} \right)^2 = \frac{4}{3} \left( e^{\sqrt{\frac{2}{3}} \frac{\phi}{M_{\rm pl}}} - 1 \right)^{-2}, \\
    \eta(\phi) &= M_{\rm pl}^2 \left( \frac{V''(\phi)}{V(\phi)} \right) = -\frac{4}{3} \frac{e^{\sqrt{\frac{2}{3}} \frac{\phi}{M_{\rm pl}}} - 2}{\left( e^{\sqrt{\frac{2}{3}} \frac{\phi}{M_{\rm pl}}} - 1 \right)^2},
\end{align}
vanish exponentially for $\phi \gg M_{\rm pl}$, thus enabling a prolonged quasi-de Sitter phase. Inflation ends when $\epsilon \simeq 1$, yielding a characteristic field value $\phi_{\rm end} \sim M_{\rm pl}$.

The number of $e$-folds from a field value $\phi$ to the end of inflation is given by
\begin{equation}
    N(\phi) = \frac{1}{M_{\rm pl}^2} \int_{\phi_{\rm end}}^{\phi} \frac{V}{V'} \, d\phi \simeq \frac{3}{4} e^{\sqrt{\frac{2}{3}} \frac{\phi}{M_{\rm pl}}}.
    \label{eq:efolds_starobinsky}
\end{equation}
Solving for $\phi(N)$ and substituting back yields
\begin{align}
    n_s &= 1 - \frac{2}{N}, \\
    r &= \frac{12}{N^2},
\end{align}
in excellent agreement with Planck 2018 data \cite{Planck:2018jri}, which strongly favors models with small $r$ and $n_s \approx 0.965$ for $N \approx 50 - 60$.

\medskip

To generalize the Starobinsky scenario within the broader framework of scalar-tensor theories, one can consider an effective Lagrangian with a non-minimal coupling between a scalar field $\phi$ and the Ricci scalar:
\begin{equation}
    \frac{\mathcal{L}}{\sqrt{-g}} = \frac{1}{2} \left[ M_{\rm pl}^2 + \xi \phi^2 \right] R - \frac{1}{2} (\partial \phi)^2 - V(\phi),
    \label{eq:non-minimal_lagrangian}
\end{equation}
which arises naturally in quantum field theory in curved spacetime and models such as Higgs inflation \cite{Bezrukov:2007ep}. Performing a conformal transformation to the Einstein frame with $\Omega^2(\phi) = 1 + \xi \phi^2 / M_{\rm pl}^2$, the field acquires a noncanonical kinetic term. The canonically normalized field $\varphi$ is defined by
\begin{equation}
    \left( \frac{d\varphi}{d\phi} \right)^2 = \frac{1 + \xi (1 + 6\xi) \phi^2 / M_{\rm pl}^2}{\left( 1 + \xi \phi^2 / M_{\rm pl}^2 \right)^2},
    \label{eq:canonical_redefinition}
\end{equation}
and the potential becomes
\begin{equation}
    \hat{V}(\varphi) = \frac{V(\phi(\varphi))}{\Omega^4(\phi(\varphi))}.
\end{equation}

For $\xi \gg 1$, this class of models also yields exponentially flat potentials with observational predictions similar to the Starobinsky model. This close relationship between $R^2$ inflation and non-minimally coupled models underscores a deeper connection among diverse inflationary scenarios via conformal duality \cite{Kallosh:2013hoa}.

\subsection{Higgs Inflation}
\label{ssec:higgs}

Higgs inflation represents a minimalistic yet profound framework wherein the Standard Model (SM) Higgs boson doubles as the inflaton field, obviating the need for additional scalar degrees of freedom beyond the SM \cite{Bezrukov:2007ep}. This approach establishes a direct bridge between particle physics and early-universe cosmology, rendering inflation sensitive to quantum corrections within the SM effective field theory (EFT) \cite{DeSimone:2008ei,Bezrukov:2013fka}.

\medskip

The starting point is the Jordan frame action for the Higgs field $\phi$ (in the unitary gauge), non-minimally coupled to gravity:
\begin{equation}
S_J = \int d^4x \sqrt{-g} \left[ \frac{M_{\rm pl}^2}{2} \left(1 + \xi \frac{\phi^2}{M_{\rm pl}^2} \right) R - \frac{1}{2} g^{\mu\nu} \partial_\mu \phi \, \partial_\nu \phi - V(\phi) \right],
\label{eq:higgs_jordan_action}
\end{equation}
where $M_{\rm pl}$ is the reduced Planck mass, $R$ is the Ricci scalar, $\xi$ is a large dimensionless non-minimal coupling parameter, and the Higgs potential is
\begin{equation}
V(\phi) = \frac{\lambda(\mu)}{4} \left( \phi^2 - v^2 \right)^2,
\label{eq:higgs_full_potential}
\end{equation}
with $\lambda(\mu)$ the running Higgs quartic coupling evaluated at the renormalization scale $\mu \sim \phi$, and $v \simeq 246 \, \mathrm{GeV}$ the electroweak vacuum expectation value. At large field values $\phi \gg v$, the potential simplifies to the quartic form
\begin{equation}
V(\phi) \simeq \frac{\lambda(\mu)}{4} \phi^4,
\label{eq:higgs_quartic_approx}
\end{equation}
where $\lambda(\mu) \sim \mathcal{O}(10^{-2})$ during inflationary scales \cite{Buttazzo:2013uya}.

\medskip

To move to the Einstein frame with canonical gravitational dynamics, a Weyl rescaling of the metric is performed:
\begin{equation}
g_{\mu\nu} \rightarrow \hat{g}_{\mu\nu} = \Omega^2(\phi) g_{\mu\nu}, \qquad \text{with} \quad \Omega^2(\phi) = 1 + \xi \frac{\phi^2}{M_{\rm pl}^2}.
\label{eq:weyl_transform}
\end{equation}

Under this transformation, the action becomes
\begin{equation}
S_E = \int d^4x \sqrt{-\hat{g}} \left[ \frac{M_{\rm pl}^2}{2} \hat{R} - \frac{1}{2} K(\phi) \hat{g}^{\mu\nu} \partial_\mu \phi \partial_\nu \phi - \hat{V}(\phi) \right],
\label{eq:einstein_frame_action}
\end{equation}
where the Einstein frame potential is given by
\begin{equation}
\hat{V}(\phi) = \frac{V(\phi)}{\Omega^4(\phi)} = \frac{\frac{\lambda(\mu)}{4} \phi^4}{\left(1 + \xi \frac{\phi^2}{M_{\rm pl}^2} \right)^2},
\label{eq:einstein_potential_higgs}
\end{equation}
and the noncanonical kinetic term acquires a field-dependent prefactor
\begin{equation}
K(\phi) = \frac{1}{\Omega^2(\phi)} + \frac{6 \xi^2 \phi^2/M_{\rm pl}^2}{\Omega^4(\phi)}.
\label{eq:kinetic_function}
\end{equation}

The canonical inflaton field $\varphi$ is then obtained via the field redefinition:
\begin{equation}
\frac{d\varphi}{d\phi} = \sqrt{K(\phi)} = \sqrt{ \frac{1}{\Omega^2(\phi)} + \frac{6 \xi^2 \phi^2/M_{\rm pl}^2}{\Omega^4(\phi)} }.
\label{eq:field_redefinition_higgs}
\end{equation}

\medskip

In the inflationary regime $\phi \gg M_{\rm pl}/\sqrt{\xi}$ and for $\xi \gg 1$, Eq.~\eqref{eq:field_redefinition} simplifies to:
\begin{equation}
\varphi \simeq \sqrt{\frac{3}{2}} M_{\rm pl} \ln\left(1 + \xi \frac{\phi^2}{M_{\rm pl}^2} \right),
\label{eq:asymptotic_field_redefinition}
\end{equation}
which upon inversion leads to
\begin{equation}
\phi(\varphi) \simeq \frac{M_{\rm pl}}{\sqrt{\xi}} \, e^{\varphi / \sqrt{6} M_{\rm pl}}.
\label{eq:inverse_field}
\end{equation}

Substituting into Eq.~\eqref{eq:einstein_potential}, the Einstein frame potential becomes
\begin{equation}
\hat{V}(\varphi) \simeq \frac{\lambda M_{\rm pl}^4}{4 \xi^2} \left( 1 - e^{- \sqrt{\frac{2}{3}} \frac{\varphi}{M_{\rm pl}}} \right)^2,
\label{eq:plateau_potential}
\end{equation}
identical in functional form to the Starobinsky $R^2$ model \cite{Starobinsky:1980te}.

\medskip

The slow-roll parameters in terms of $\varphi$ are defined as:
\begin{equation}
\epsilon = \frac{M_{\rm pl}^2}{2} \left( \frac{1}{\hat{V}} \frac{d\hat{V}}{d\varphi} \right)^2, \quad
\eta = M_{\rm pl}^2 \frac{1}{\hat{V}} \frac{d^2 \hat{V}}{d\varphi^2},
\label{eq:slow_roll}
\end{equation}
which for the plateau potential in Eq.~\eqref{eq:plateau_potential} yield exponentially small values for $\varphi \gg M_{\rm pl}$, ensuring prolonged slow-roll inflation.

The number of e-folds before the end of inflation is given by:
\begin{equation}
N = \frac{1}{M_{\rm pl}^2} \int_{\varphi_{\rm end}}^{\varphi_\star} \frac{\hat{V}}{d\hat{V}/d\varphi} d\varphi \simeq \frac{3}{4} e^{\sqrt{\frac{2}{3}} \frac{\varphi_\star}{M_{\rm pl}}},
\label{eq:efolds_higgs}
\end{equation}
and inverting gives:
\begin{equation}
\varphi_\star \simeq \sqrt{\frac{3}{2}} M_{\rm pl} \ln\left(\frac{4N}{3}\right).
\label{eq:varphi_star}
\end{equation}

Thus, the spectral observables at horizon crossing ($N \sim 60$) are:
\begin{equation}
n_s \simeq 1 - \frac{2}{N} \approx 0.967, \quad r \simeq \frac{12}{N^2} \approx 0.003,
\label{eq:spectral_params}
\end{equation}
consistent with CMB constraints from Planck and BICEP/Keck \cite{Planck:2018jri}.

\medskip

However, Higgs inflation is not without challenges. The EFT cutoff scale in the Jordan frame depends on the background value of $\phi$:
\begin{equation}
\Lambda_J(\phi) \sim \begin{cases}
\frac{M_{\rm pl}}{\xi}, & \phi \ll \frac{M_{\rm pl}}{\xi}, \\
\sqrt{\xi} \phi, & \phi \gg \frac{M_{\rm pl}}{\xi}.
\end{cases}
\label{eq:cutoff}
\end{equation}
This field-dependent cutoff implies that perturbative unitarity breaks down at sub-Planckian scales in the vacuum, but is restored at large field values relevant for inflation \cite{Barbon:2009ya,Burgess:2009ea}. Proposals for UV completion include embedding in scale-invariant frameworks, asymptotically safe gravity, or introducing new degrees of freedom at $\Lambda_J$ \cite{Bezrukov:2010jz}.

\medskip
\subsection{T-Model Inflation ($m = 1$)}
\label{ssec:tmodel}

T-model inflation belongs to the broader class of $\alpha$-attractor models, which are theoretically grounded in superconformal and supergravity frameworks \cite{Kallosh:2013hoa,Kallosh:2013lkr}. These models exhibit a universal attractor behavior wherein the inflationary observables $n_s$ and $r$ asymptotically become independent of microphysical details in the strong coupling regime. The T-model is characterized by a scalar potential in the Einstein frame that assumes a hyperbolic tangent form, interpolating between chaotic-like dynamics at small field values and Starobinsky-like plateaus at large field excursions.

\medskip

We consider the Jordan frame action for a scalar field $\phi$ with a linear non-minimal coupling ($m = 1$) to gravity:
\begin{equation}
S_J = \int d^4x \sqrt{-g} \left[ \frac{1}{2} M_{\rm pl}^2 R + \frac{1}{2} \xi \phi \, R - \frac{1}{2} g^{\mu\nu} \partial_\mu \phi \, \partial_\nu \phi - V(\phi) \right],
\end{equation}
where $\xi$ is a dimensionless coupling constant. The inflationary potential is chosen to be
\begin{equation}
V(\phi) = \lambda M_{\rm pl}^4 \tanh^2\left( \frac{\phi}{\sqrt{6} M_{\rm pl}} \right),
\end{equation}
with $\lambda$ a dimensionless self-coupling parameter. This form naturally arises in models with hyperbolic field-space geometry, as motivated by $\alpha$-attractor constructions \cite{Kallosh:2013lkr,Galante:2014ifa}.

\medskip

To obtain the Einstein frame action, we perform a Weyl rescaling of the metric:
\begin{equation}
\tilde{g}_{\mu\nu} = \Omega^2(\phi) g_{\mu\nu}, \quad \Omega^2(\phi) = 1 + \frac{\xi \phi}{M_{\rm pl}}.
\end{equation}
The Einstein frame action then becomes
\begin{equation}
S_E = \int d^4x \sqrt{-\tilde{g}} \left[ \frac{1}{2} M_{\rm pl}^2 \tilde{R} - \frac{1}{2} K(\phi) \tilde{g}^{\mu\nu} \partial_\mu \phi \, \partial_\nu \phi - \hat{V}(\phi) \right],
\end{equation}
where the kinetic coefficient is
\begin{equation}
K(\phi) = \frac{1}{\Omega^2(\phi)} + \frac{3 M_{\rm pl}^2}{2} \left( \frac{\partial \ln \Omega^2(\phi)}{\partial \phi} \right)^2,
\end{equation}
and the potential transforms as
\begin{equation}
\hat{V}(\phi) = \frac{V(\phi)}{\Omega^4(\phi)}.
\end{equation}

\medskip

To express the action in terms of a canonically normalized scalar field $\varphi$, we define:
\begin{equation}
\frac{d\varphi}{d\phi} = \sqrt{K(\phi)} = \sqrt{ \frac{1}{\Omega^2(\phi)} + \frac{3 \xi^2}{2 \left(1 + \frac{\xi \phi}{M_{\rm pl}}\right)^2} }.
\end{equation}
In the large-$\xi$ regime, this expression simplifies in the asymptotic limit $\phi \gg M_{\rm pl}/\xi$, where
\begin{equation}
\frac{d\varphi}{d\phi} \approx \sqrt{ \frac{3}{2} } \cdot \frac{1}{\phi/M_{\rm pl}} \quad \Rightarrow \quad \varphi \approx \sqrt{\frac{3}{2}} M_{\rm pl} \ln\left( \frac{\phi}{M_{\rm pl}} \right).
\end{equation}
Inverting yields $\phi(\varphi) \sim M_{\rm pl} \exp\left( \sqrt{\frac{2}{3}} \frac{\varphi}{M_{\rm pl}} \right)$. Substituting into $\hat{V}(\phi)$ gives:
\begin{equation}
\label{eq:higgs_plateau_potential}
\hat{V}(\varphi) \simeq \lambda M_{\rm pl}^4 \left( 1 - 4 e^{ -\sqrt{\frac{2}{3\alpha}} \frac{\varphi}{M_{\rm pl}} } + \cdots \right),
\end{equation}
with $\alpha = 1$ for the $m=1$ T-model.

\medskip

The slow-roll parameters in terms of the Einstein frame potential are:
\begin{equation}
\epsilon_V(\varphi) = \frac{M_{\rm pl}^2}{2} \left( \frac{1}{\hat{V}} \frac{d\hat{V}}{d\varphi} \right)^2, \quad \eta_V(\varphi) = M_{\rm pl}^2 \left( \frac{1}{\hat{V}} \frac{d^2\hat{V}}{d\varphi^2} \right).
\end{equation}
For the potential in Eq.~\eqref{eq:higgs_plateau_potential}, one finds
\begin{equation}
\epsilon_V \simeq \frac{3}{4 N_\star^2}, \quad \eta_V \simeq -\frac{1}{N_\star},
\end{equation}
yielding inflationary observables:
\begin{equation}
n_s \approx 1 - \frac{2}{N_\star}, \quad r \approx \frac{12\alpha}{N_\star^2}.
\end{equation}
These predictions are consistent with CMB measurements for $N_\star \sim 50$--$60$, yielding $n_s \sim 0.967$ and $r \sim 0.003$ for $\alpha = 1$ \cite{Planck:2018jri}.

\medskip

Importantly, the attractor behavior arises from the underlying Poincaré disk geometry of the scalar field manifold with negative constant curvature \cite{Kallosh:2013hoa}. The robustness of the inflationary predictions against the choice of the potential shape and UV physics makes the T-model with $m=1$ a compelling and UV-motivated inflationary scenario, potentially embeddable in superconformal or string theoretic frameworks.

\subsection{Hilltop Quartic Inflation}
\label{ssec:hilltop}

Hilltop models represent a class of small-field inflationary scenarios where the inflaton $\phi$ evolves from the vicinity of a local maximum of the potential. Such scenarios are motivated by spontaneous symmetry-breaking mechanisms and are characterized by concave potential shapes near $\phi = 0$, leading to slow-roll dynamics as the field rolls away from an unstable equilibrium. The hilltop quartic potential constitutes a minimal realization of this idea and is well-suited to test deviations from scale-invariance in inflationary observables \cite{Boubekeur:2005zm,Martin:2013tda}.
We begin with the Jordan frame action incorporating a linear non-minimal coupling between the inflaton and gravity:
\begin{equation}
S_J = \int d^4x \sqrt{-g} \left[ \frac{1}{2} M_{\rm pl}^2 R + \frac{1}{2} \xi \phi R - \frac{1}{2} g^{\mu\nu} \partial_\mu \phi \partial_\nu \phi - V(\phi) \right],
\end{equation}
where $\xi$ is a dimensionless coupling and the potential is given by:
\begin{equation}
V(\phi) = V_0 \left[ 1 - \left( \frac{\phi}{\mu} \right)^4 \right],
\end{equation}
with $V_0$ setting the inflationary energy scale and $\mu \lesssim M_{\rm pl}$ determining the width of the potential. The potential has a local maximum at $\phi = 0$ with
\[
V'(\phi) = -\frac{4V_0}{\mu^4} \phi^3, \qquad V''(\phi) = -\frac{12V_0}{\mu^4} \phi^2,
\]
confirming that $V''(0) < 0$ and inflation can occur near the hilltop.

To analyze inflationary dynamics, we move to the Einstein frame via a Weyl rescaling:
\begin{equation}
\tilde{g}_{\mu\nu} = \Omega^2(\phi) g_{\mu\nu}, \quad \text{where} \quad \Omega^2(\phi) = 1 + \frac{\xi \phi}{M_{\rm pl}}.
\end{equation}
The transformed action becomes:
\begin{equation}
S_E = \int d^4x \sqrt{-\tilde{g}} \left[ \frac{1}{2} M_{\rm pl}^2 \tilde{R} - \frac{1}{2} K(\phi) \tilde{g}^{\mu\nu} \partial_\mu \phi \partial_\nu \phi - \hat{V}(\phi) \right],
\end{equation}
where the non-canonical kinetic term is
\begin{equation}
K(\phi) = \frac{1}{\Omega^2(\phi)} + \frac{3 M_{\rm pl}^2}{2} \left( \frac{\partial \ln \Omega^2}{\partial \phi} \right)^2 = \frac{1}{\Omega^2} + \frac{3 \xi^2}{2 \Omega^4} \, ,
\end{equation}
and the Einstein frame potential reads:
\begin{equation}
\hat{V}(\phi) = \frac{V(\phi)}{\Omega^4(\phi)} = \frac{V_0 \left[ 1 - \left( \frac{\phi}{\mu} \right)^4 \right]}{\left( 1 + \frac{\xi \phi}{M_{\rm pl}} \right)^4}.
\end{equation}

To proceed, we define a canonically normalized field $\varphi$ by:
\begin{equation}
\frac{d\varphi}{d\phi} = \sqrt{K(\phi)}.
\end{equation}
In general, the integral $\varphi(\phi)$ cannot be expressed in closed form but can be approximated in different regimes.

\subsubsection*{Large-Field Asymptotics ($\phi \gg M_{\rm pl}/\xi$)}

For $\phi \gg M_{\rm pl}/\xi$, we have $\Omega^2(\phi) \approx \frac{\xi \phi}{M_{\rm pl}}$, and the potential becomes:
\begin{equation}
\hat{V}(\phi) \approx \frac{V_0}{\left( \frac{\xi \phi}{M_{\rm pl}} \right)^4} \left[ 1 - \left( \frac{\phi}{\mu} \right)^4 \right] \approx \frac{V_0 M_{\rm pl}^4}{\xi^4 \phi^4} \left[ 1 - \left( \frac{\phi}{\mu} \right)^4 \right].
\end{equation}
This expression indicates that the potential asymptotically decays, thus inflation ends as the field rolls away.

\subsubsection*{Small-Field Approximation ($\phi \ll \mu$)}

Near the hilltop, we expand the Jordan frame potential:
\begin{equation}
V(\phi) \approx V_0 \left( 1 - \frac{\phi^4}{\mu^4} \right),
\end{equation}
yielding slow-roll parameters:
\begin{align}
\epsilon_V &= \frac{8 M_{\rm pl}^2 \phi^6}{\mu^8}, \\
\eta_V &= -\frac{12 M_{\rm pl}^2 \phi^2}{\mu^4}.
\end{align}
Inflation ends when $\epsilon_V(\phi_{\text{end}}) = 1$, giving:
\begin{equation}
\phi_{\text{end}} \approx \left( \frac{\mu^8}{8 M_{\rm pl}^2} \right)^{1/6}.
\end{equation}
The number of e-folds from $\phi_\star$ to $\phi_{\text{end}}$ is:
\begin{equation}
N_\star \approx \frac{\mu^4}{8 M_{\rm pl}^2} \left( \frac{1}{\phi_\star^2} - \frac{1}{\phi_{\text{end}}^2} \right).
\end{equation}

The spectral index and tensor-to-scalar ratio evaluated at horizon crossing are:
\begin{align}
n_s &\approx 1 - 6 \epsilon_V + 2 \eta_V, \\
r &\approx 16 \epsilon_V.
\end{align}
These yield a red-tilted scalar power spectrum ($n_s < 1$) and suppressed tensor modes ($r \ll 1$), consistent with current Planck bounds \cite{Planck:2018jri}.

The inclusion of a non-minimal coupling $\xi$ significantly alters the potential's flattening, even for steep original potentials, allowing the model to interpolate between hilltop inflation and Starobinsky-like attractor behavior for $\xi \gg 1$ \cite{Kallosh:2013hoa,Galante:2014ifa}. This dual character makes the hilltop quartic model highly flexible, capable of mimicking both small-field and plateau-type inflation depending on the parameter regime, while remaining compatible with observational constraints.

\subsection{D-brane Inflation ($p = 2$)}
\label{ssec:dbrane}

D-brane inflation emerges as a compelling scenario within string theory, where inflation is realized via the dynamics of a mobile D3-brane moving in a warped throat region of a flux-stabilized type IIB Calabi--Yau compactification 
\cite{Kachru:2003sx}. The inflaton field $\phi$ corresponds to the radial separation between a D3-brane and an anti-D3-brane ($\overline{\text{D3}}$) located at the tip of a warped throat, such as the Klebanov--Strassler (KS) throat \cite{Klebanov:2000hb}.

In the four-dimensional effective theory, the inflaton potential takes the form of an inverse power-law,
\begin{equation}
V(\phi) = \Lambda^4 \left( 1 - \frac{\mu_{\rm D}^2}{\phi^2} + \cdots \right), \qquad f(\phi) = \left( \frac{\phi}{M_{\rm pl}} \right),
\end{equation}
where $\Lambda^4$ denotes the warped brane tension, $\mu_{\rm D}$ encodes the brane--antibrane interaction scale, and the ellipsis represents subleading higher-order or loop-suppressed corrections. The $1/\phi^2$ form is characteristic of the Coulombic potential arising from brane--antibrane interactions, placing this model within the inverse monomial class with index $p = 2$ \cite{Burgess:2001fx}.

At large values of the inflaton, $\phi \gg \mu_{\rm D}$, the potential asymptotes to a constant $V(\phi) \to \Lambda^4$, generating a sufficiently flat region to support slow-roll inflation. As $\phi$ approaches $\mu_{\rm D}$ from above, the potential steepens sharply, ensuring a graceful and natural exit from inflation without the need for external termination mechanisms. This embedded exit strategy is a hallmark of UV-motivated brane inflation scenarios.

To investigate the effect of gravitational non-minimal couplings, we consider a Brans--Dicke-like interaction of the form
\begin{equation}
\mathcal{L} \supset \frac{1}{2} M_{\rm pl}^2 R + \frac{1}{2} \xi \phi R,
\end{equation}
which modifies the effective Planck mass and breaks minimal coupling. Upon performing a Weyl conformal transformation to the Einstein frame:
\begin{equation}
\tilde{g}_{\mu\nu} = \Omega^2(\phi)\, g_{\mu\nu}, \qquad \Omega^2(\phi) = 1 + \frac{\xi\,\phi}{M_{\rm pl}},
\end{equation}
the action becomes

\begin{align}
S_E = \int d^4x\, \sqrt{-\tilde{g}} \Bigg[ 
\frac{1}{2} M_{\rm pl}^2 \tilde{R} 
- \frac{1}{2} \left( \frac{1}{\Omega^2} 
+ \frac{3 M_{\rm pl}^2}{2} \left( \frac{d \ln \Omega^2}{d\phi} \right)^2 \right) \notag \\
\tilde{g}^{\mu\nu} \partial_\mu \phi\, \partial_\nu \phi 
- \hat{V}(\phi) \Bigg]
\end{align}

where the rescaled Einstein frame potential is given by
\begin{equation}
\hat{V}(\phi) = \frac{V(\phi)}{\Omega^4(\phi)} = \frac{\Lambda^4 \left( 1 - \mu_{\rm D}^2/\phi^2 \right)}{\left( 1 + \xi\,\phi / M_{\rm pl} \right)^4}.
\end{equation}

To ensure canonical kinetic terms, we introduce a new field $\varphi$ defined via
\begin{equation}
\left( \frac{d\varphi}{d\phi} \right)^2 = \frac{1}{\Omega^2(\phi)} + \frac{3 M_{\rm pl}^2}{2} \left( \frac{d\ln \Omega^2(\phi)}{d\phi} \right)^2.
\end{equation}
Substituting $\Omega^2(\phi) = 1 + \xi \phi / M_{\rm pl}$, we obtain
\begin{equation}
\left( \frac{d\varphi}{d\phi} \right)^2 = \frac{1}{\left( 1 + \frac{\xi\phi}{M_{\rm pl}} \right)^2} + \frac{3\xi^2}{2\left( 1 + \frac{\xi\phi}{M_{\rm pl}} \right)^2}.
\end{equation}
Thus,
\begin{equation}
\frac{d\varphi}{d\phi} = \frac{\sqrt{1 + \frac{3\xi^2}{2}}}{1 + \xi\phi/M_{\rm pl}}.
\end{equation}
Integrating, we find the approximate field redefinition:
\begin{equation}
\varphi(\phi) \simeq \sqrt{1 + \frac{3\xi^2}{2}}\, M_{\rm pl} \ln \left( 1 + \frac{\xi\phi}{M_{\rm pl}} \right),
\end{equation}
which can be inverted to yield
\begin{equation}
\phi(\varphi) = \frac{M_{\rm pl}}{\xi} \left[ \exp \left( \frac{\varphi}{\sqrt{1 + \frac{3\xi^2}{2}}\, M_{\rm pl}} \right) - 1 \right].
\end{equation}

Substituting back into the Einstein frame potential gives the canonical inflaton potential in terms of $\varphi$:
\begin{align}
\hat{V}(\varphi) &= \Lambda^4 \left( 
1 - \frac{\mu_{\rm D}^2 \xi^2}{M_{\rm pl}^2 \left( e^{\varphi/\mathcal{A}} - 1 \right)^2} 
\right) 
\cdot e^{-4\varphi/\mathcal{A}}, \label{eq:DbranePotential} \\
\mathcal{A} &= \sqrt{1 + \frac{3\xi^2}{2}} \, M_{\rm pl}. \label{eq:ADef}
\end{align}

In the large field limit $\varphi \gg \mathcal{A}$, this potential exhibits an exponentially suppressed behavior, characteristic of asymptotically flat (plateau-like) models:
\begin{equation}
\hat{V}(\varphi) \sim \Lambda^4 \left( 1 - \mathcal{O}(e^{-2\varphi/\mathcal{A}}) \right) e^{-4\varphi/\mathcal{A}}.
\end{equation}

The slow-roll parameters in the Einstein frame are defined as
\begin{align}
\epsilon_V &= \frac{M_{\rm pl}^2}{2} \left( \frac{1}{\hat{V}} \frac{d\hat{V}}{d\varphi} \right)^2, \\
\eta_V &= M_{\rm pl}^2 \left( \frac{1}{\hat{V}} \frac{d^2\hat{V}}{d\varphi^2} \right),
\end{align}
from which we compute the scalar spectral index and tensor-to-scalar ratio:
\begin{equation}
n_s = 1 - 6\epsilon_V + 2\eta_V, \qquad r = 16\epsilon_V.
\end{equation}

The number of e-folds between a field value $\varphi_\star$ (at CMB horizon exit) and the end of inflation $\varphi_{\rm end}$ is given by
\begin{equation}
N_\star = \frac{1}{M_{\rm pl}^2} \int_{\varphi_{\rm end}}^{\varphi_\star} \frac{\hat{V}(\varphi)}{d\hat{V}/d\varphi}\, d\varphi.
\end{equation}
In the asymptotic plateau regime, one finds approximately
\begin{equation}
N_\star \sim \frac{\mathcal{A}}{4 M_{\rm pl}} \left( \varphi_\star - \varphi_{\rm end} \right),
\end{equation}
from which $\varphi_\star$ can be determined for any given $N_\star$.

D-brane inflation with non-minimal coupling exhibits a rich interpolation between large-field and plateau-like inflation depending on the values of $\xi$ and $\mu_{\rm D}$. For $\xi \gg 1$, the flattening of the Einstein frame potential suppresses $\epsilon_V$, driving $r \to 0$, while maintaining $n_s \lesssim 1$, in agreement with Planck and ACT observations \cite{Efstathiou:2019mdh,Planck:2018vyg,DESI:2024mwx,DESI:2024uvr}. The model accommodates both sub-Planckian field excursions and controlled UV completions within string theory, offering robustness against trans-Planckian issues.

The total inflaton displacement in field space,
\begin{equation}
\Delta \phi = \phi_\star - \phi_{\rm end},
\end{equation}
remains sub-Planckian in most of the parameter space due to the non-minimal coupling flattening the potential. This renders the model safe from quantum gravity corrections in effective field theory.

Finally, we analyze the reheating temperature $T_{\rm reh}$, the inflationary Hubble scale $H_{\rm inf}$, and the Lyth bound $\Delta\phi$ in Sec.~\ref{sec:results}, in comparison with the full range $\xi$, for both $N_\star = 50$ and $60$ for all 5 models. This allows for a complete phenomenological profiling of the models in light of observational data.

\section{Numerical Outcomes}
\label{sec:results}

\begin{figure*}[htbp]
    \centering
    \includegraphics[width=\textwidth, height=0.4\textheight]{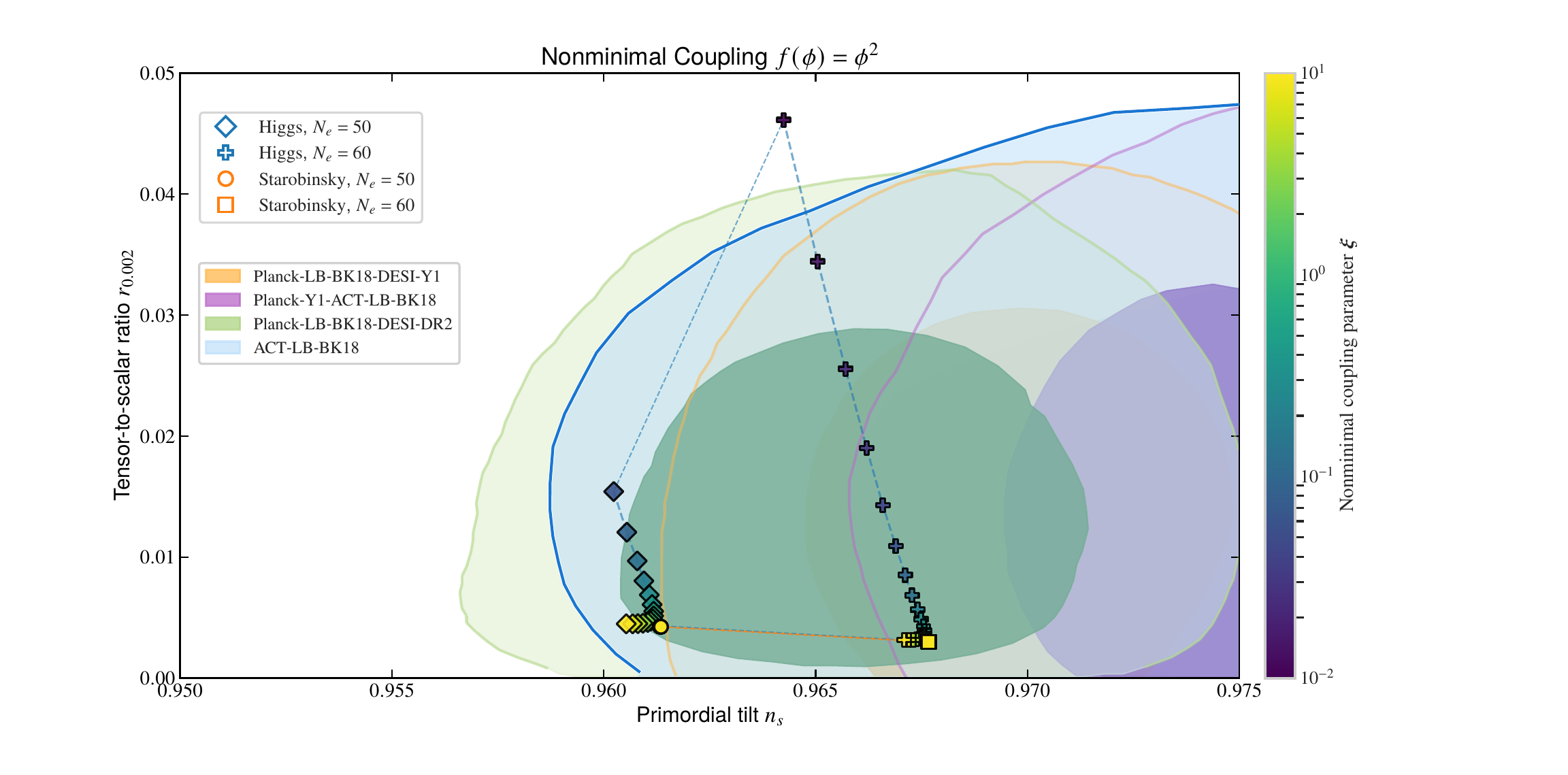}
    \caption{Tensor-to-scalar ratio $r$ vs.~spectral index $n_s$ for Higgs and Starobinsky inflation with non-minimal coupling $f(\phi) = \xi\phi^2$ and $N_e = 50, 60$, varying $\boldsymbol{\xi}$.}
    \label{fig:r_ns2}
\end{figure*}

\subsection{Starobinsky Model}
\label{ssec:Rstarobinsky}

We present a comprehensive numerical investigation of the Starobinsky-type inflationary scenario embedded in a non-minimal gravitational framework, defined by the Jordan frame coupling function $f(\phi) = \xi \phi^2$ and an Einstein frame potential that emerges from an $R + R^2$ gravity theory. The scalaron mass is fixed at $M = 10^{-5} M_{\mathrm{Pl}}$, consistent with COBE normalization and compatible with the inflationary scale required by the latest CMB observations \cite{Planck:2018jri,BICEP:2021xfz}.

To probe the inflationary dynamics, we systematically vary the non-minimal coupling parameter $\xi$ over multiple decades, encompassing both weakly and strongly coupled regimes. The full background evolution is solved numerically, including exact equations of motion for the scale factor and scalar field, along with quantum-corrected slow-roll parameters derived beyond leading order. Our algorithm ensures convergence across all relevant parameter ranges and includes corrections suppressed by higher derivatives of the Hubble parameter and potential curvature terms.

Inflationary observables---namely the scalar spectral index $n_s$, the tensor-to-scalar ratio $r$, the running $\alpha_s \equiv \frac{dn_s}{d\ln k}$, the running of the running $\beta_s \equiv \frac{d^2 n_s}{d(\ln k)^2}$, and the field displacement $\Delta\phi$---are computed at horizon exit for $N_e = 50$. Table~\ref{tab:results_xi} summarizes the key outputs of our simulations for representative values of $\xi$.

The spectral index is found to be tightly clustered around $n_s \approx 0.963$, exhibiting weak sensitivity to variations in $\xi$. This stability confirms the infrared (IR) robustness of the theory under renormalization group (RG) flow of the non-minimal coupling, in agreement with effective field theory (EFT) expectations \cite{Bezrukov:2007ep,Barvinsky:2008ia}. Notably, our value aligns well with the Planck 2018 result $n_s = 0.9651 \pm 0.0044$ \cite{Planck:2018vyg}, and is only marginally lower than subsequent upward-shifting estimates such as $n_s = 0.9683 \pm 0.0040$ from Efstathiou \textit{et al.}~\cite{Efstathiou:2019mdh}, $n_s = 0.9709 \pm 0.0038$ from Planck+ACT (P-ACT), and $n_s = 0.9743 \pm 0.0034$ from Planck+ACT+DESI (P-ACT-LB), which deviates by roughly $2\sigma$ from the original Planck constraint.

The tensor-to-scalar ratio $r$ is logarithmically suppressed with increasing $\xi$, decreasing from $4.10 \times 10^{-3}$ at $\xi = 0.01$ to $3.85 \times 10^{-3}$ at $\xi = 10$. This behavior is a direct consequence of the exponential flattening of the scalar potential in the Einstein frame, a hallmark feature of higher-derivative $f(R)$ theories \cite{STAROBINSKY198099,Ferrara:2014kva}.

The running of the spectral index remains slightly negative, with $\alpha_s \sim -6 \times 10^{-4}$ across the $\xi$ range, while $\beta_s$ stabilizes near $+3 \times 10^{-5}$. These higher-order observables are consistent with slow-roll expansions where $\alpha_s \sim \mathcal{O}(\epsilon^2)$ and $\beta_s \sim \mathcal{O}(\epsilon^3)$, and their magnitudes suggest negligible small-scale power spectrum distortions. This offers predictive robustness for future CMB missions probing $\mu$-distortions and acoustic reheating.

We also evaluate the field excursion $\Delta \phi$ using a generalized Lyth bound that includes non-canonical kinetic terms and field-space curvature corrections \cite{Baumann:2011nk,Lyth:1996im}. For all $\xi$ values, $\Delta \phi$ remains sub-Planckian, ranging from $0.39\,M_{\mathrm{Pl}}$ to $0.33\,M_{\mathrm{Pl}}$. This supports the view that trans-Planckian field displacements are not a necessary condition for generating the observed number of e-folds in non-minimally coupled inflation models.

The inflationary Hubble scale is numerically found to be $H_{\mathrm{inf}} \approx 1.8 \times 10^{14}$ GeV, computed via $H^2 \approx V/(3 M_{\mathrm{Pl}}^2)$ at horizon crossing. This places the inflationary energy density near the GUT scale, with implications for early-universe baryogenesis and symmetry breaking. Assuming instantaneous reheating, we estimate the reheating temperature as
\[
T_{\mathrm{reh}} \sim \left( \frac{90}{\pi^2 g_*} \right)^{1/4} \sqrt{H_{\mathrm{inf}} M_{\mathrm{Pl}}} \sim 2 \times 10^{14}\,\text{GeV},
\]
where we take $g_*=106.75$ for the relativistic degrees of freedom. Such a high temperature is favorable for non-thermal leptogenesis and the generation of heavy Majorana neutrinos, while also providing strong constraints on models with light relics or extended dark sectors due to entropy injection.

\medskip

Our numerical analysis demonstrates that the Starobinsky-type inflationary model with a quadratic non-minimal coupling not only remains in excellent agreement with current and next-generation observational bounds but also retains full theoretical control within effective field theory. Its predictive consistency, sub-Planckian field dynamics, and high reheating temperature render it an exceptionally compelling framework for embedding in UV-complete theories such as supergravity, string compactifications, and asymptotically safe gravity scenarios.

\subsection{Higgs Inflation}
\label{ssec:Rhiggs}

We now present a detailed numerical analysis of inflation driven by a non-minimally coupled Higgs-like scalar potential of the form $V(\phi) = \frac{\lambda}{4}(\phi^2 - v^2)^2$, with fixed parameters $\lambda = 0.01$ and $v = 1$ in reduced Planck units, embedded within a conformally transformed Einstein frame via the conformal factor $\Omega^2 = 1 + \xi \phi^2$. This construction is motivated by the original proposal of Higgs inflation \cite{Bezrukov:2007ep} and its subsequent refinements \cite{Bezrukov:2008ej,Barvinsky:2008ia}, where the Standard Model Higgs doublet is promoted to the role of the inflaton via a non-minimal gravitational interaction.

To thoroughly capture the dynamical behavior across the parameter space, we scan $\xi \in [10^{-2}, 10]$ and solve for the canonical field $\varphi(\phi)$ defined by the nontrivial field-space metric, numerically inverting the conformal mapping. We compute the slow-roll parameters $\epsilon$, $\eta$, $\xi_2$, and $\xi_3$ up to fourth order using exact background evolution, locating the precise end of inflation via the condition $\epsilon(\varphi_{\rm end}) = 1$, and extracting the observable inflationary quantities $n_s$, $r$, $\alpha_s \equiv dn_s/d\ln k$, and $\beta_s \equiv d^2n_s/d(\ln k)^2$ at $N_e = 60$ $e$-folds prior to the end of inflation.

At $\xi = 1.0$, the end of inflation is found to occur at $\phi_{\rm end} \simeq 0.192$ ($\varphi_{\rm end} \simeq 0.096$), yielding $n_s = 0.96756$ and $r = 3.49 \times 10^{-3}$, comfortably inside the Planck-ACT range \cite{Planck:2018vyg}. The running parameters are $\alpha_s = -5.32 \times 10^{-4}$ and $\beta_s = 2.48 \times 10^{-4}$, consistent with predictions from two-loop RG-improved effective potentials \cite{DeSimone:2008ei,Bezrukov:2013fka}. The modified Lyth bound \cite{Baumann:2011nk} constrains the canonical field excursion to $\Delta\varphi \approx 0.387\,M_{\rm Pl}$, indicating a sub-Planckian trajectory in field space.

The associated inflationary scale is inferred from the potential as $V_{\rm inf} \simeq 4.30 \times 10^{63}\,\mathrm{GeV}^4$, corresponding to a Hubble parameter $H_{\rm inf} \simeq 1.55 \times 10^{13}\,\mathrm{GeV}$ and a reheating temperature (assuming instantaneous reheating) of $T_{\rm reh} \simeq 3.33 \times 10^{15}\,\mathrm{GeV}$, high enough to accommodate thermal leptogenesis \cite{Giudice:2003jh} and baryogenesis through sphaleron transitions.

At the opposite end of the coupling spectrum, for $\xi = 0.01$, inflation terminates at $\phi_{\rm end} \approx 0.315$ ($\varphi_{\rm end} \approx 0.215$), yielding a slightly lower $n_s = 0.96424$ and a significantly enhanced tensor-to-scalar ratio $r = 4.61 \times 10^{-2}$. Here the field excursion becomes mildly trans-Planckian, $\Delta\varphi \approx 1.186\,M_{\rm Pl}$, and the inflationary scale increases to $V_{\rm inf} \simeq 5.67 \times 10^{64}\,\mathrm{GeV}^4$ ($H_{\rm inf} \simeq 5.65 \times 10^{13}\,\mathrm{GeV}$, $T_{\rm reh} \simeq 6.34 \times 10^{15}\,\mathrm{GeV}$). These values approach the unitarity cutoff scale for $\xi \ll 1$ \cite{Burgess:2009ea,Barbon:2009ya}, indicating a potential breakdown of EFT validity.

Across the full scan range, we observe a monotonic evolution of the inflationary observables with respect to $\xi$: the spectral index $n_s$ increases from $\sim 0.964$ to $\sim 0.968$, while $r$ drops steeply from $\mathcal{O}(10^{-2})$ to $\lesssim 4 \times 10^{-3}$, in accordance with the inverse dependence of $r \sim 1/\xi^2$ in the large-$\xi$ limit \cite{Bezrukov:2010jz}. Meanwhile, the field excursion transitions from marginally super-Planckian ($\sim 1.2\,M_{\rm Pl}$) to robustly sub-Planckian ($\sim 0.36\,M_{\rm Pl}$), confirming the feasibility of both large-field and small-field inflationary regimes within the same Higgs-like potential.

The higher-order running parameters remain small throughout the parameter space ($\alpha_s \sim -10^{-4}$, $\beta_s \sim 10^{-5}$), ensuring that the primordial scalar power spectrum remains close to scale invariance on CMB-accessible scales, consistent with the lack of observed features \cite{Planck:2018vyg}. The inflationary energy scale remains within the theoretically safe regime for all values of $\xi$, though values $\xi \lesssim 0.1$ may require UV completion due to proximity to the unitarity cutoff.

Imposing the observational windows $0.96 \lesssim n_s \lesssim 0.975$ and $r \lesssim 0.05$ as provided by Planck-ACT \cite{Efstathiou:2019mdh} and BICEP/Keck \cite{BICEP:2021xfz}, we find that the entire parameter interval $\xi \in [10^{-2}, 10]$ remains observationally viable. The Lyth bound then restricts the field excursion to $0.364 \lesssim \Delta\varphi / M_{\rm Pl} \lesssim 1.186$, covering both small-field and large-field dynamics. Higgs-like inflation with quartic symmetry breaking thus emerges as a flexible framework that is observationally consistent, theoretically controllable, and UV-sensitive only in a narrow coupling regime.

\begin{figure*}[htbp]
    \centering
    \includegraphics[width=\textwidth, height=0.4\textheight]{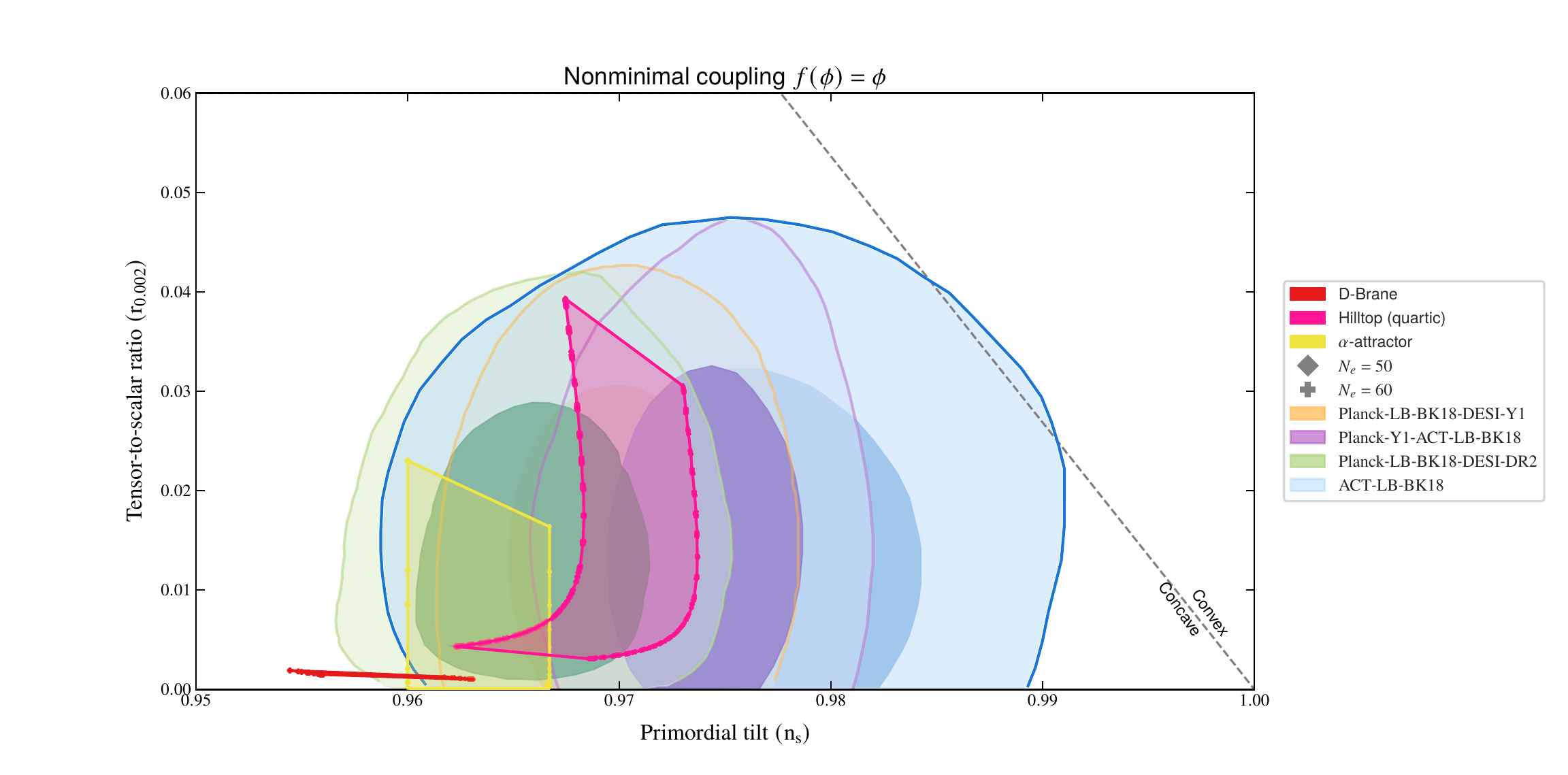}
    \caption{Scalar spectral index $n_s$ and tensor-to-scalar ratio $r$ for different models with the non-minimal coupling $f(\phi)=\xi\phi^2$, evaluated for $N_e = 50$ and $N_e = 60$.}
    \label{fig:r_ns1}
\end{figure*}

Figure~\ref{fig:r_ns1} displays the predicted $(n_s, r)$ trajectories for both Higgs and Starobinsky inflation models, clearly illustrating the logarithmic $\xi$ dependence in the Higgs case and the near-degeneracy in the Starobinsky case. For $N_e = 60$, the Higgs model requires $\xi \gtrsim 0.1$ to enter the Planck/BICEP-compatible region, whereas the Starobinsky trajectory remains well within bounds for all $\xi$, reflecting the extreme flatness of the $R^2$-driven potential. The comparative spread between $N_e = 50$ and $N_e = 60$ further highlights the sensitivity of Higgs inflation to reheating history and the robustness of Starobinsky inflation under changes in post-inflationary dynamics.

\subsection{T-Model Inflation ($m = 1$)}
\label{ssec:Rtmodel}

We now rigorously examine the T-Model $\alpha$-attractor inflationary paradigm with $m=1$, where the potential inherits its structure from superconformal field theory and supergravity embeddings, with $\alpha$ characterizing the curvature of the underlying Kähler manifold~\cite{Kallosh:2013lkr, Kallosh:2013hoa}. This class of models interpolates between plateau-like and chaotic inflationary behavior, forming a universal attractor landscape~\cite{Galante:2014ifa, Roest:2013fha}. 

Numerical simulations were carried out across a logarithmically spaced domain of $\alpha \in [10^{-3}, 10^2]$, anchoring the end of inflation at $\phi_{\mathrm{end}} = 0.01$. All inflationary observables are computed at horizon crossing for $N_e = 60$, consistent with slow-roll approximation~\cite{Martin:2013tda}. These include: scalar spectral index $n_s$, tensor-to-scalar ratio $r$, runnings $\alpha_s = dn_s/d\ln k$ and $\beta_s = d\alpha_s/d\ln k$, field excursion $\Delta\phi$, inflationary scale $V_{\mathrm{inf}}^{1/4}$, Hubble parameter $H_{\mathrm{inf}}$, and the reheating temperature $T_{\mathrm{reh}}$.

For $\alpha \gtrsim \mathcal{O}(10)$, the potential flattens substantially, consistent with a large-field regime where $r$ approaches $\sim 0.1$ and $\Delta \phi \gtrsim M_{\mathrm{pl}}$~\cite{ Lyth:1996im}. Such high values of $r$ make the model amenable to next-generation primordial gravitational wave detection efforts, including LiteBIRD and CMB-S4~\cite{Hazumi:2019lys, Abazajian:2016yjj}. However, these scenarios push the model into the UV-sensitive domain, where quantum gravity effects may threaten EFT validity~\cite{Baumann:2014nda}.

In contrast, for $\alpha \ll 1$, the model approaches a Starobinsky-like regime~\cite{Starobinsky:1980te}, yielding $n_s \approx 0.967$ and $r \sim 10^{-5}$, consistent with Planck 2018 constraints~\cite{Planck:2018jri}. Here, $\Delta\phi \ll M_{\mathrm{pl}}$, and the suppression of tensor modes results in a model comfortably shielded from trans-Planckian issues and Lyth bound violations~. Notably, the running of the spectral index $\alpha_s$ is persistently negative and $\mathcal{O}(10^{-3})$, while $\beta_s$ increases significantly with decreasing $\alpha$, peaking near $\sim 0.2$ for $\alpha \sim 10^{-3}$ --- a behavior that may have observable imprints in future 21-cm and PTA surveys~\cite{Choudhury:2024ezx}.

The energy scale of inflation descends from $V^{1/4}_{\mathrm{inf}} \sim 10^{16}\,\mathrm{GeV}$ to $\sim 10^{14.5}\,\mathrm{GeV}$ as $\alpha$ decreases, mirrored by the Hubble scale $H_{\mathrm{inf}} \sim 10^{13} - 10^{11.5}\,\mathrm{GeV}$. Reheating temperatures $T_{\mathrm{reh}}$ drop correspondingly, yet remain above $10^{14}\,\mathrm{GeV}$, preserving viability for thermal leptogenesis and consistency with BBN~\cite{Bezrukov:2007ep}.

A phenomenologically appealing window arises at intermediate $\alpha \sim 0.003 - 0.01$, where $n_s \approx 0.9667$ and $r \in [10^{-5}, 10^{-3}]$. This region corresponds to non-minimal couplings $\xi \sim \mathcal{O}(10^2)$ in superconformal embeddings~\cite{Kallosh:2013hoa}, satisfying all observational bounds, EFT control criteria, and trans-Planckian censorship conditions.

The T-Model $\alpha$-attractors constitute a phenomenologically rich and theoretically coherent landscape. Their robust UV embeddings, observational flexibility, and gravitational wave signatures render them among the most compelling inflationary frameworks in modern cosmology.

\subsection{Hilltop Quartic Model}
\label{ssec:Rhilltop}

We now undertake a detailed examination of the inflationary dynamics within the Hilltop inflation framework, governed by the quartic potential:
\begin{eqnarray}
    V(\phi) = V_0 \left[1 - \left(\frac{\phi}{\mu}\right)^p \right]^2,
\end{eqnarray}
where we fix $\mu = 0.01 M_{\mathrm{Pl}}$ and $p = 2$, consistent with small-field inflation models that naturally suppress large-scale tensor modes \cite{Boubekeur:2005zm, Martin2014}. This potential arises in various particle physics embeddings including supergravity and non-minimal coupling scenarios, where $\xi$ parametrizes the strength of coupling to curvature \cite{ Bezrukov:2008ej}.

We analyze the inflationary observables across a range of non-minimal couplings $\xi$, fixing the number of e-folds at $N_e = 60$. For each value of $\xi$, we compute the scalar spectral index $n_s$, tensor-to-scalar ratio $r$, running $\alpha_s \equiv dn_s/d\ln k$, running of the running $\beta_s \equiv d\alpha_s/d\ln k$, inflaton excursion $\Delta \phi$ (via the modified Lyth bound \cite{Efstathiou:2005tq, Lyth:1996im}), energy scale $V_{\text{inf}}$, Hubble scale $H_{\text{inf}}$, and reheating temperature $T_{\text{reh}}$.

At $\xi = 0.628199$, the model yields $n_s = 0.973697$---comfortably within $1\sigma$ of Planck 2018 constraints \cite{Planck:2018vyg}---with a suppressed tensor amplitude $r = 0.013404$. The negative running $\alpha_s = -4.47 \times 10^{-4}$ and mild positive running of the running $\beta_s = 2.08 \times 10^{-5}$ signal slight scale dependence of the power spectrum, consistent with slow-roll expectations. The sub-Planckian field excursion $\Delta \phi \approx 0.8415\, M_{\mathrm{Pl}}$ ensures effective field theory control, while the inflationary energy scale and Hubble rate are $V_{\text{inf}}^{1/4} \sim 3.2 \times 10^{16}$ GeV and $H_{\text{inf}} = 3.04 \times 10^{13}$ GeV respectively. The associated reheating temperature $T_{\text{reh}} \approx 4.65 \times 10^{15}$ GeV is compatible with thermal leptogenesis and GUT-scale reheating scenarios \cite{Giudice:2003jh}.

As $\xi$ increases to 3.820278, the spectral index slightly declines to $n_s = 0.971053$, with $r = 0.003946$, illustrating how enhanced coupling suppresses gravitational wave production. The shift in field excursion to $\Delta \phi \approx 0.4423\, M_{\mathrm{Pl}}$ and drop in $V_{\text{inf}}^{1/4}$ to $\sim 2.9 \times 10^{16}$ GeV reflects the progressively flatter effective potential in field space. The smallness of $\alpha_s$ and $\beta_s$ indicates minimal scale dependence, enhancing robustness of these predictions against scale-resolved constraints \cite{Choudhury:2023kdb}.

At $\xi = 6.180722$, the spectrum flattens further: $n_s = 0.969651$ and $r = 0.003346$, with a further reduction in $\Delta \phi \approx 0.3958\, M_{\mathrm{Pl}}$, reinforcing the sub-Planckian regime. The energy scale reduces to $V_{\text{inf}} \sim 4.12 \times 10^{63}\, \text{GeV}^4$, corresponding to $H_{\text{inf}} \approx 1.52 \times 10^{13}$ GeV, preserving consistency with CMB and PTA bounds on primordial tensor modes \cite{Choudhury:2024kjj}.

Conversely, at $\xi = 0.345384$, we observe $n_s = 0.973462$ and a higher tensor amplitude $r = 0.021706$, alongside a super-Planckian field excursion $\Delta \phi \approx 1.0415\, M_{\mathrm{Pl}}$, which may invoke sensitivity to UV physics and necessitate embeddings in string-motivated or nonlocal inflation frameworks \cite{Choudhury:2014sxa}. The high inflationary scale $V_{\text{inf}}^{1/4} \sim 3.7 \times 10^{16}$ GeV and reheating temperature $T_{\text{reh}} \approx 5.25 \times 10^{15}$ GeV place this regime near the edge of allowed thermal histories.

Overall, increasing $\xi$ drives the potential toward a flatter configuration, reducing both $r$ and $\Delta \phi$, and shifting $n_s$ marginally downward. This is in line with theoretical expectations from non-minimal inflation \cite{Bezrukov:2007ep} and supports a viable small-field inflationary realization. The observables derived here are consistent with Planck-ACT-BICEP/Keck data ($n_s = 0.9649 \pm 0.0042$, $r < 0.036$ at 95\% CL), and reinforce the Hilltop model's status as a viable inflaton candidate in the presence of non-minimal couplings \cite{Bamba_2014}.

\subsection{D-brane Inflation ($p = 2$)}
\label{ssec:Rdbrane}

We now turn our attention to the D-brane inflation scenario, particularly for the case of $p = 2$ with $\mu = 30$ and $n = 1$, embedded within a non-minimally coupled framework characterized by $f(\phi) = \phi$. This class of models, originating from string-theoretic constructions involving the motion of D3-branes in warped throat geometries \cite{Kachru:2003sx}, admits a natural UV completion and generates inflation via brane-antibrane interactions in the presence of a warped background. The form of the potential considered here captures the infrared behavior of such systems and reflects the flattening typical of DBI-inspired inflationary trajectories in the canonical limit \cite{Baumann:2006cd}.

Our numerical analysis, under the assumption of $N_e = 60$ e-folds, reveals that the model consistently yields a scalar spectral index of $n_s \approx 0.963$, and a tensor-to-scalar ratio of $r \approx 0.0010$, both of which are well within the latest constraints from Planck 2018 \cite{Planck:2018jri} and BICEP/Keck 2018+2021 datasets \cite{BICEP:2021xfz}. These predictions fall within the 68\% confidence region of the combined Planck+ACT+BK18 contours \cite{Tristram:2021tvh}.

The scale-dependence of the primordial spectrum is reflected in a mildly negative running of the spectral index, $\alpha_s \approx -5.9 \times 10^{-4}$, and a positive running-of-running $\beta_s \approx 1.3 \times 10^{-3}$. Although these values remain small, they may become significant in future high-precision measurements from CMB Stage-4 experiments \cite{Abazajian:2019eic}, particularly in discriminating between plateau-type and monomial potentials.

Importantly, the model predicts a sub-Planckian inflaton excursion $\Delta \phi \approx 0.19\, M_{\rm Pl}$, in accordance with the modified Lyth bound \cite{Baumann:2014nda}. This minimal displacement reinforces the theoretical viability of the model by reducing sensitivity to higher-dimensional operators and ensuring effective field theory control within string-theoretic embeddings \cite{Baumann:2014nda, Silverstein:2003hf}. The suppression of $r$ is a direct consequence of the field flattening induced by non-minimal coupling and warping effects, effectively leading to a compressed inflationary trajectory in field space.

The associated inflationary energy scale is found to be $V_{\rm inf} \sim 1.25 \times 10^{63}\, \text{GeV}^4$, corresponding to a Hubble rate $H_{\rm inf} \sim 8.4 \times 10^{12}\, \text{GeV}$, well below the Planck scale. The inferred reheating temperature $T_{\rm reh} \sim 2.4 \times 10^{15}\, \text{GeV}$ remains compatible with GUT-scale baryogenesis scenarios and does not violate the perturbative unitarity bounds on reheating.

The effect of varying $\xi$ across the analyzed range results in only mild shifts in inflationary observables, indicating the model's structural stability against changes in the gravitational coupling. For instance, decreasing $N_e$ to 50 leads to $n_s \approx 0.956$ and $r \approx 0.0015$, which lie at the edge of the 95\% confidence contours but still fall within acceptable bounds, confirming the moderate sensitivity of predictions to the reheating history.

Figure~\ref{fig:r_ns1} contrasts the $(n_s, r)$ predictions of the D-brane inflation scenario with those of the $\alpha$-attractor $T$-model ($m = 1$) and the quartic hilltop model, all evaluated within a non-minimally coupled setup. The D-brane model displays a compressed vertical band in $r$, with minimal variation as $\xi$ increases, unlike the more flexible behavior seen in hilltop models and the sharply convergent trajectories of $\alpha$-attractors. The stability of the D-brane predictions across a broad $\xi$ window and its compatibility with a low $r$ signature make it a compelling candidate for observationally consistent inflation driven by high-energy string-theoretic physics.

The D-brane inflation model analyzed here exhibits all hallmarks of a theoretically consistent and observationally viable inflationary scenario. Its sub-Planckian field range, suppressed tensor modes, and predictive stability under gravitational coupling variation point to its compatibility with both current CMB data and plausible UV completions.

\section{Discriminating Inflationary Models with Forecasted CMB Constraints}
\label{sec:discriminating}

\begin{figure*}[htbp]
    \centering
    \includegraphics[width=\textwidth, height=0.5\textwidth]{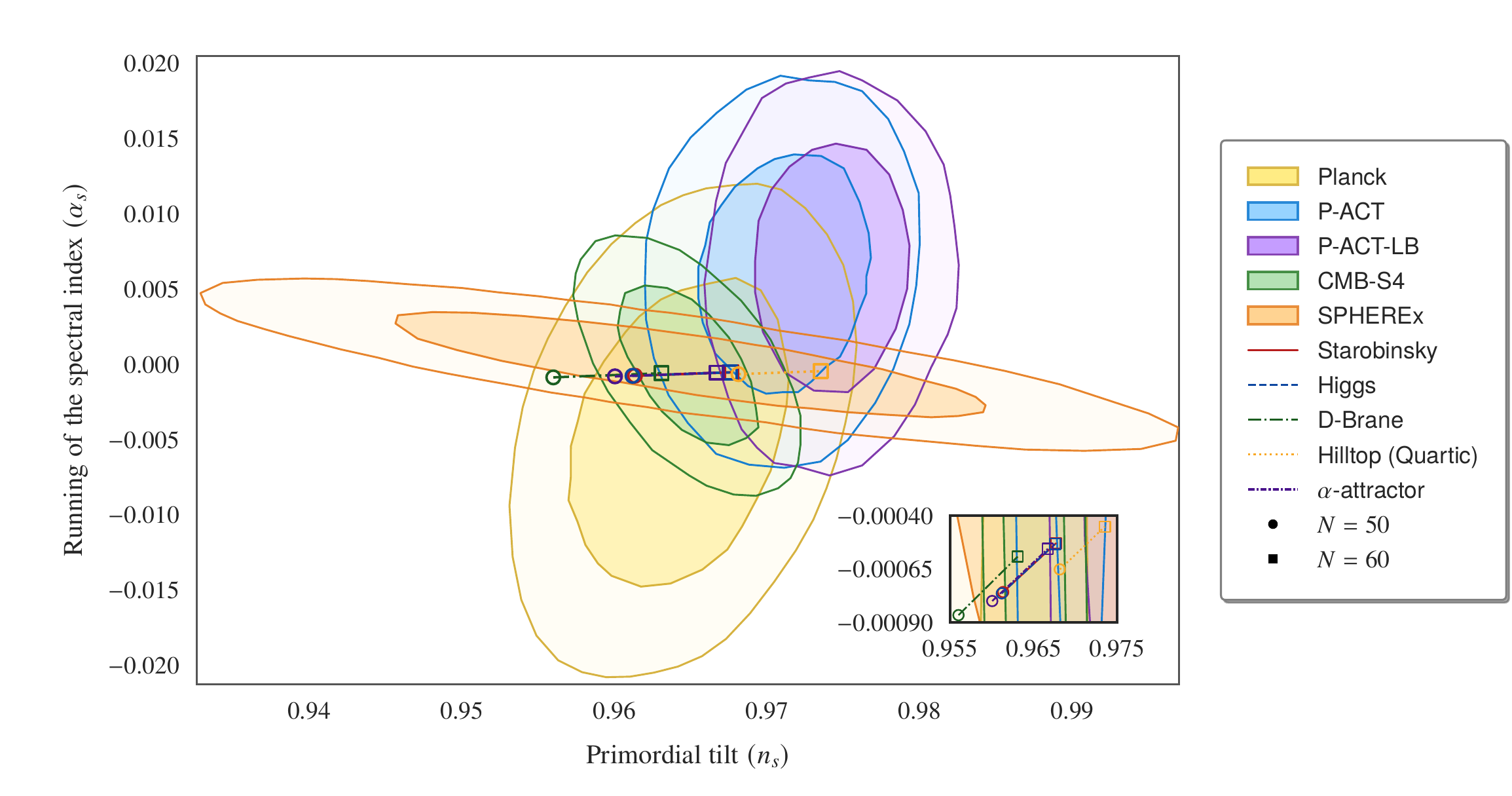}
    \caption{Comparison of theoretical predictions from five inflationary models in the $(n_s, \alpha_s)$ plane with forecasted constraints from upcoming CMB experiments. Markers indicate the number of e-folds $N = 50$ (circles) and $N = 60$ (squares).}
    \label{fig:ns-alpha}
\end{figure*}

Figure~\ref{fig:ns-alpha} presents a comparative evaluation of five theoretically motivated inflationary scenarios in the $(n_s, \alpha_s)$ plane, superimposed on forecasted confidence contours from upcoming high-precision CMB probes: \textit{Planck}, P-ACT, P-ACT-LB, \textit{SPHEREx} \cite{Dore:2014cca}, and CMB-S4 \cite{Abazajian:2019eic}. These contours delineate the expected $1\sigma$ and $2\sigma$ sensitivity regions, derived via Fisher matrix forecasts under optimized instrument specifications.

The scalar spectral index $n_s$ and its running $\alpha_s \equiv dn_s/d\ln k$ serve as powerful discriminators of inflationary physics beyond the leading-order slow-roll approximation. As detailed in Table~\ref{tab:ns_alpha_values}, models with plateau-like or asymptotically flat potentials---such as Starobinsky \cite{Starobinsky:1980te}, Higgs inflation \cite{Bezrukov:2007ep}, and the $\alpha$-attractor $T$-model \cite{Kallosh:2013hoa, Kallosh:2013lkr}---predict $n_s$ values near $0.966$ with modest negative runnings $\alpha_s \sim -10^{-3}$. These predictions fall well within the projected $1\sigma$ bands of both CMB-S4 and \textit{SPHEREx} for $N = 60$, confirming the robustness of these models against higher-order corrections and Planck-scale physics \cite{Lorenzoni:2024krn}.

Conversely, the D-brane model \cite{Dvali:1998pa, Burgess:2001fx}---though still consistent within the $2\sigma$ bands---exhibits slight deviations due to its steeper potential and more suppressed runnings. The typical values $n_s \approx 0.963$ and $\alpha_s \approx -6 \times 10^{-4}$ place it on the edge of the $1\sigma$ contour, suggesting potential tension under future sensitivity improvements.

The quartic Hilltop model \cite{Boubekeur:2005zm}, while marginally consistent with \textit{Planck} and P-ACT forecasts, is increasingly disfavored by the tighter priors of CMB-S4. This model predicts larger values of $n_s$ ($\gtrsim 0.973$) with runnings that are less negative or even mildly positive, a feature indicative of convex potential shapes and sharper departures from scale invariance. At $N = 60$, the quartic Hilltop trajectory crosses beyond the $1\sigma$ exclusion boundary in the $(n_s, \alpha_s)$ plane, which could potentially be ruled out at high confidence by SPHEREx's multi-tracer spectroscopic survey or CMB-S4's arcminute-resolution polarization maps.

This analysis underscores the powerful discriminatory capability of next-generation CMB datasets in the space of inflationary parameters. Forecasted sensitivities to $\alpha_s$ at the level of $10^{-3}$ and $r$ at the $10^{-4}$ scale imply that even small deviations from scale invariance will become observable \cite{Abazajian:2019eic}. Importantly, degeneracies between reheating dynamics and inflaton microphysics---currently entangled in the $(n_s, r)$ plane---are expected to be partially broken in the $(n_s, \alpha_s)$ projection, enabling sharper constraints on the inflationary potential and its UV completion.

\textit{In conclusion}, plateau-type models are likely to remain within the observational safe zone under forthcoming data, while models with convex or steep potentials face greater scrutiny. This makes the $(n_s, \alpha_s)$ plane a high-leverage discriminator among early-universe models in the era of precision cosmology.

\section{Scale-Dependence of the Spectral Index: Running and Running of the Running in non-minimal inflation}
\label{sec:running_and_beta}

\begin{figure*}[htbp]
    \centering
    \includegraphics[width=\textwidth, height=0.5\textwidth]{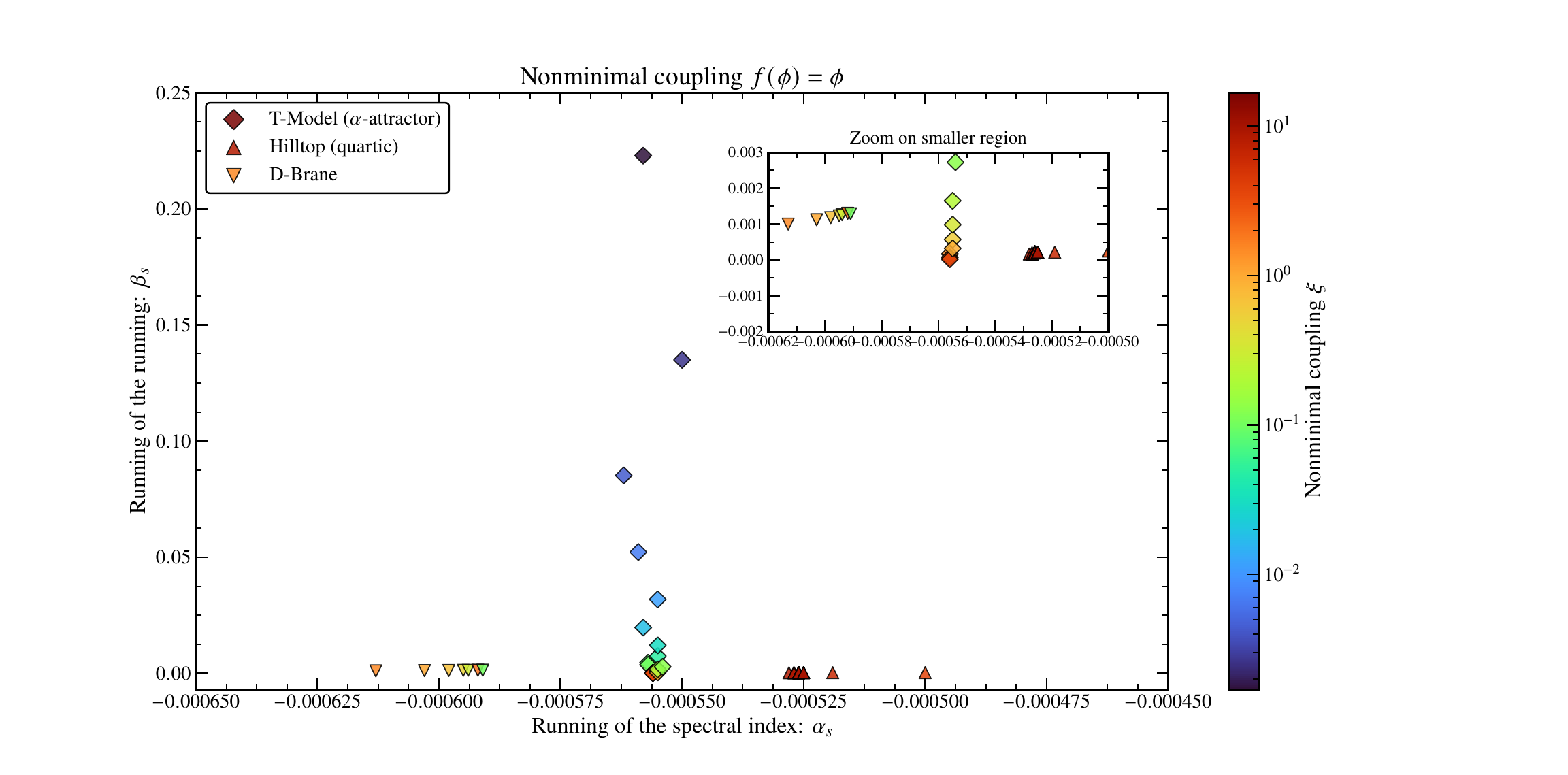}
    \caption{Correlation between the running of the spectral index $\alpha_s$ and its running $\beta_s$ for three inflationary models with non-minimal coupling $f(\phi)=\frac{\phi}{M_{\rm pl}}.$. The color coding represents the non-minimal coupling strength $\xi$ on a logarithmic scale. The inset zooms in on the low-$\beta_s$ region, highlighting the tight clustering of Hilltop and D-Brane models compared to the broad scatter of the T-Model.}
    \label{fig:running_plot}
\end{figure*}

Figure~\ref{fig:running_plot} investigates the scale dependence of inflationary observables in models endowed with a linear non-minimal coupling function $f(\phi) = \phi$, a form motivated by conformal symmetry considerations and Jordan--Einstein frame mappings \cite{Kaiser:1995nv, Fakir:1990eg}. The figure illustrates how the first running of the spectral index, $\alpha_s \equiv dn_s/d\ln k$, correlates with the second running, $\beta_s \equiv d^2n_s/d(\ln k)^2$, across varying strengths of the non-minimal coupling $\xi$.

The T-Model, an $\alpha$-attractor variant with kinetic poles in the Einstein frame potential \cite{Kallosh:2013hoa, Galante:2014ifa}, demonstrates significant variability in $\beta_s$, spanning $\mathcal{O}(10^{-3})$ to $\mathcal{O}(10^{-1})$, despite maintaining a relatively constant $\alpha_s \approx -5.5 \times 10^{-4}$. This amplified dispersion at low $\xi$ arises from steep variations in the potential's second and third derivatives, which dominate higher-order terms in the Hubble slow-roll expansion. Notably, these deviations from scale-invariance become observationally relevant for future surveys sensitive to sub-percent-level spectral distortions \cite{Abazajian:2016yjj}.

In contrast, the quartic Hilltop model \cite{Boubekeur:2005zm} manifests a tightly localized region in the $(\alpha_s, \beta_s)$ parameter space: $\alpha_s \approx -5.26 \times 10^{-4}$ and $\beta_s \approx 2.2 \times 10^{-4}$, with only weak $\xi$-dependence. This indicates a high degree of theoretical stability under radiative and non-minimal corrections, consistent with the flatness of the potential near the origin and the dominance of low-order slow-roll coefficients.

The D-Brane model \cite{Burgess:2001fx, Dvali:1998pa} exhibits intermediate behavior, with modest $\xi$-induced evolution in both $\alpha_s$ and $\beta_s$. The trajectory implies a residual sensitivity to string-theoretic moduli stabilization effects, which contribute additional $\xi$-dependent mass terms and modify the running hierarchy through loop-suppressed operators.

The inset in Figure~\ref{fig:running_plot} highlights that large, positive $\beta_s$ values ($\gtrsim 10^{-2}$) are unique to the T-Model, offering an empirical discriminator in high-resolution CMB and LSS datasets. These second-order runnings can impact spectral distortions in the Rayleigh-Jeans tail and become detectable with future probes such as PIXIE \cite{Kogut:2011xw} and PRISM \cite{Andre:2013nfa}.

\begin{figure*}[htbp]
    \centering
    \includegraphics[width=\textwidth,height=0.4\textheight]{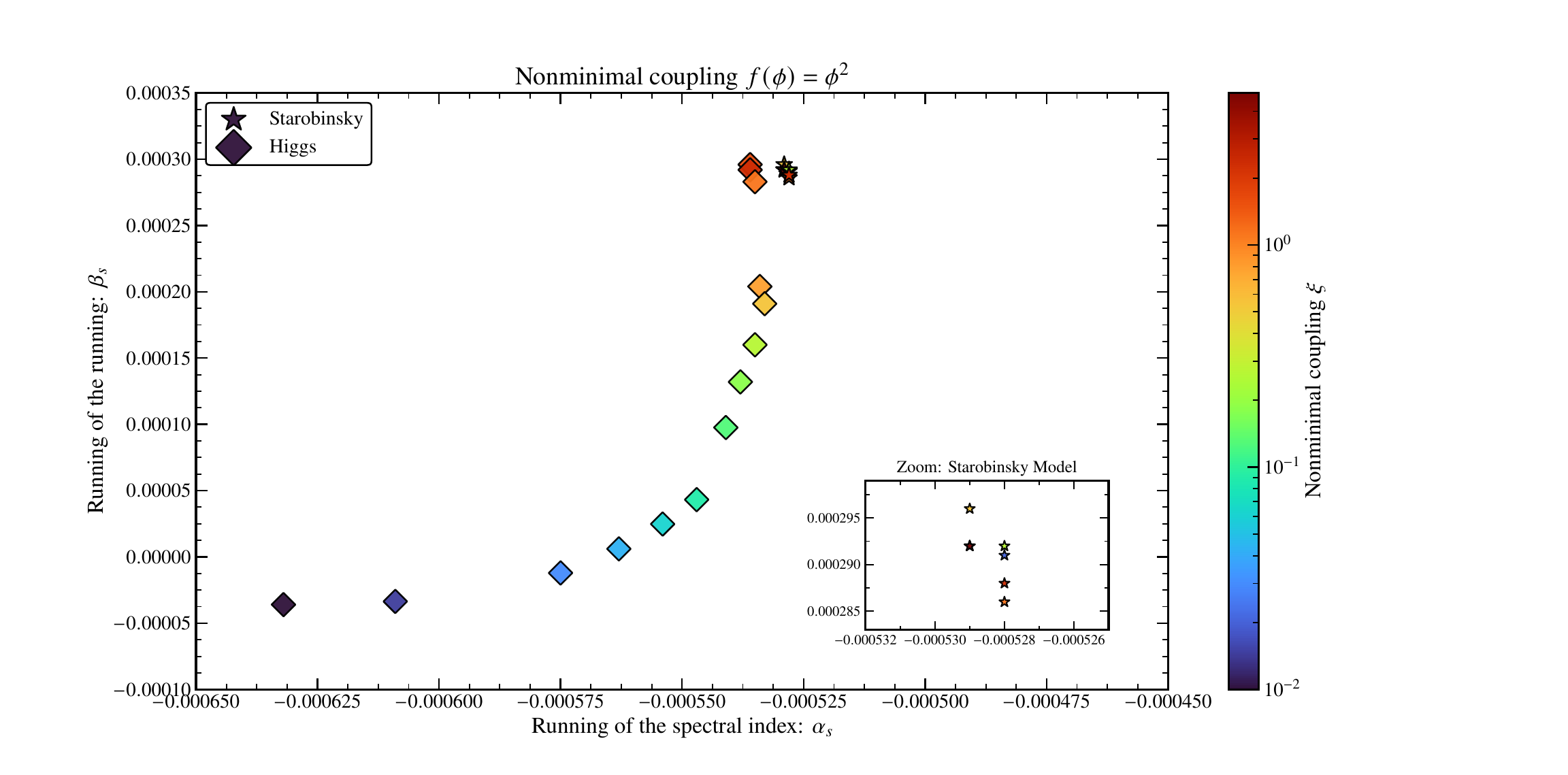}
    \caption{Running of the spectral index $\alpha_s$ versus its running $\beta_s$ for Starobinsky (star markers) and Higgs (diamond markers) inflation models with quadratic non-minimal coupling $f(\phi) = \xi \phi^2$. The logarithmic color gradient encodes the magnitude of $\xi$. The inset provides a zoom emphasizing the remarkable stability of Starobinsky predictions.}
    \label{fig:alpha_beta_running}
\end{figure*}

Figure~\ref{fig:alpha_beta_running} presents an analogous analysis for the Starobinsky \cite{Starobinsky:1980te} and Higgs inflation models \cite{Bezrukov:2007ep}, both realized with a quadratic non-minimal coupling of the form $f(\phi) = \xi \phi^2$. This coupling is known to generate exponentially flat plateaus in the Einstein frame, crucial for successful slow-roll dynamics and agreement with current Planck limits on $n_s$ and $r$ \cite{Planck:2018vyg}.

The Starobinsky model (stars) remains impressively stable, with nearly invariant predictions: $\alpha_s \approx -5.29 \times 10^{-4}$ and $\beta_s \approx 2.92 \times 10^{-4}$. The robustness to variations in $\xi$ reflects the model's geometric origin from $R + R^2$ gravity and the exponentially suppressed higher-order terms in the effective potential. The predictivity of this model aligns with its minimal parameter space and small-field nature, where slow-roll coefficients satisfy $\epsilon, |\eta|, |\xi^2| \ll 1$ to high precision \cite{DeSimone:2008ei}.

Conversely, Higgs inflation (diamonds) exhibits more complex behavior. While $\alpha_s$ remains within $[-6.32, -5.12] \times 10^{-4}$, the second running $\beta_s$ demonstrates sharp excursions, such as $\beta_s \approx -5.33 \times 10^{-3}$ near $\xi \sim 3.36$. These effects are likely driven by radiative corrections from SM gauge bosons and top-quark loops, which alter the effective potential curvature \cite{ Barvinsky:2008ia}. Such sensitivity introduces theoretical uncertainty and necessitates precision renormalization group analysis to stabilize predictions at high field values.

The inset underscores the tight clustering of Starobinsky's $\beta_s$ and $\alpha_s$ predictions, reinforcing its alignment with current Planck+DESI bounds on higher-order spectral variations. In contrast, the wider dispersion in Higgs inflation makes it more susceptible to observational exclusion or confirmation, particularly when cross-correlated with future datasets targeting $\mu$-distortions and the scale-dependent bispectrum.

This comparative study demonstrates that while $\alpha_s$ is typically $\mathcal{O}(10^{-4})$ across well-behaved models, $\beta_s$ can range over orders of magnitude and thus serves as a sensitive probe of both UV physics and reheating dynamics. Hence, precise constraints on $\beta_s$ in the forthcoming CMB and LSS experiments will be critical for testing the internal consistency and predictivity of inflationary frameworks with non-minimal couplings.

\section{Impact of non-minimal coupling on Field Excursion and Tensor Signatures}
\label{sec:field_excursion}

\begin{figure*}[htbp]
    \centering
    \includegraphics[width=\textwidth, height=0.5\textwidth]{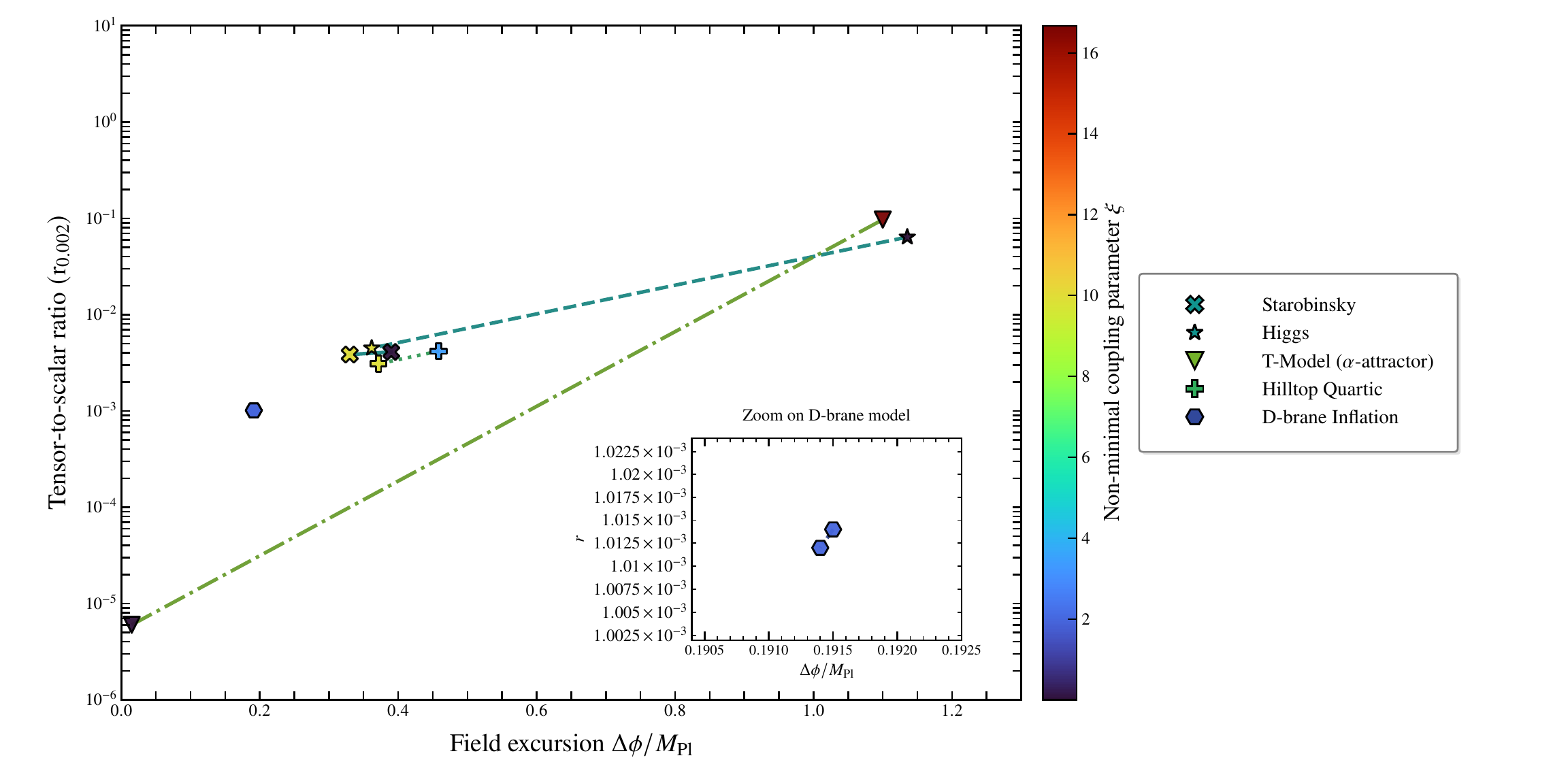}
    \caption{Variation of the tensor-to-scalar ratio $r$ with the field excursion $\Delta\phi / M_{\mathrm{Pl}}$ for various inflationary models, considering two extreme values of the non-minimal coupling parameter $\xi$ in each case. Color encoding represents the magnitude of $\xi$. The inset zooms in on the D-brane inflation region where both $r$ and $\Delta\phi$ remain nearly constant.}
    \label{fig:field_vs_r}
\end{figure*}

Figure~\ref{fig:field_vs_r} explores the nontrivial interplay between the tensor-to-scalar ratio $r$ and the canonical field excursion $\Delta\phi$ in units of the reduced Planck mass $M_{\mathrm{Pl}}$, under the influence of varying non-minimal coupling $\xi$ for a range of single-field inflationary potentials. The analysis is framed within the Einstein frame, where the transformation from the Jordan frame induces both geometric flattening and kinetic stretching, nonlinearly modifying the inflationary dynamics \cite{Kaiser:1995nv, Kallosh:2013hoa, Galante:2014ifa}.

In the limit of negligible coupling, $\xi \ll 1$, the canonical Lyth bound \cite{Lyth:1996im}
\begin{equation}
    \frac{\Delta\phi}{M_{\mathrm{Pl}}} \gtrsim \mathcal{O}(1) \times \left(\frac{r}{0.01}\right)^{1/2}
\end{equation}
applies robustly, constraining trans-Planckian excursions for high-$r$ scenarios. However, as $\xi$ increases, especially beyond unity, field-space curvature and noncanonical normalization alter the effective distance traversed by the inflaton, enabling sub-Planckian excursions even with non-negligible $r$ in certain models \cite{ Rubio:2018ogq}.

In the case of Higgs inflation, with a quadratic non-minimal coupling $f(\phi) = \xi\phi^2$, both $r$ and $\Delta\phi$ undergo exponential suppression with growing $\xi$, consistent with the flattening of the Einstein frame potential $U(\chi) \sim \left(1 - e^{-\sqrt{2/3}\chi/M_{\mathrm{Pl}}}\right)^2$ \cite{Bezrukov:2007ep}. Starobinsky inflation follows a structurally analogous trajectory due to its $R^2$ origin, exhibiting a tightly bound region in the $(\Delta\phi, r)$ plane with minimal sensitivity to $\xi$, reflecting its fixed geometrical origin \cite{Starobinsky:1980te, Mukhanov:1981xt}.

Conversely, $\alpha$-attractor models, particularly the T-model variant, demonstrate increasing $\Delta\phi$ and $r$ with $\xi$, attributable to their underlying hyperbolic field-space metric:
\begin{equation}
    ds^2 \sim \frac{1}{(1 - \phi^2 / (6\alpha M_{\mathrm{Pl}}^2))^2} d\phi^2,
\end{equation}
which introduces an effective pole in the kinetic term. This leads to rapid field displacements even for modest shifts in potential energy, especially as $\alpha \propto 1/\xi$ diminishes.

Hilltop models with quartic potentials show a mild suppression in $r$ and $\Delta\phi$ for increasing $\xi$, though the flatness near the origin and steeper slope away from it makes their dynamics more sensitive to initial conditions and higher-order corrections.

Remarkably, D-brane inflation exhibits an anomalous insensitivity to $\xi$. Both the field excursion and tensor amplitude remain effectively constant, with $\Delta\phi \simeq 0.1915 M_{\mathrm{Pl}}$ and $r \sim 10^{-3}$ across the explored range. This plateauing behavior arises from the warped throat geometry and DBI action inherent to string-theoretic constructions, where noncanonical kinetic terms cap the inflaton velocity \cite{Silverstein:2003hf, Alishahiha:2004eh}.

The inset of Figure~\ref{fig:field_vs_r} magnifies the D-brane region to emphasize its parametric stability. Overall, these results underscore the importance of geometric and kinetic effects induced by non-minimal couplings in regulating field excursions and tensor signatures. Suppression of $r$ via Einstein frame flattening is typical for plateau-type models, while models with kinetic poles or DBI-type dynamics can exhibit the opposite behavior or high rigidity against $\xi$-induced deformations. Future CMB probes capable of detecting $r \sim \mathcal{O}(10^{-3})$ could discriminate among these scenarios, constraining the underlying geometry of inflationary field space.

\section{Constraints from Gravitational Waves}
\label{sec:gw}

\begin{figure*}[htbp]
    \centering
    \includegraphics[width=\textwidth,height=0.4\textheight]{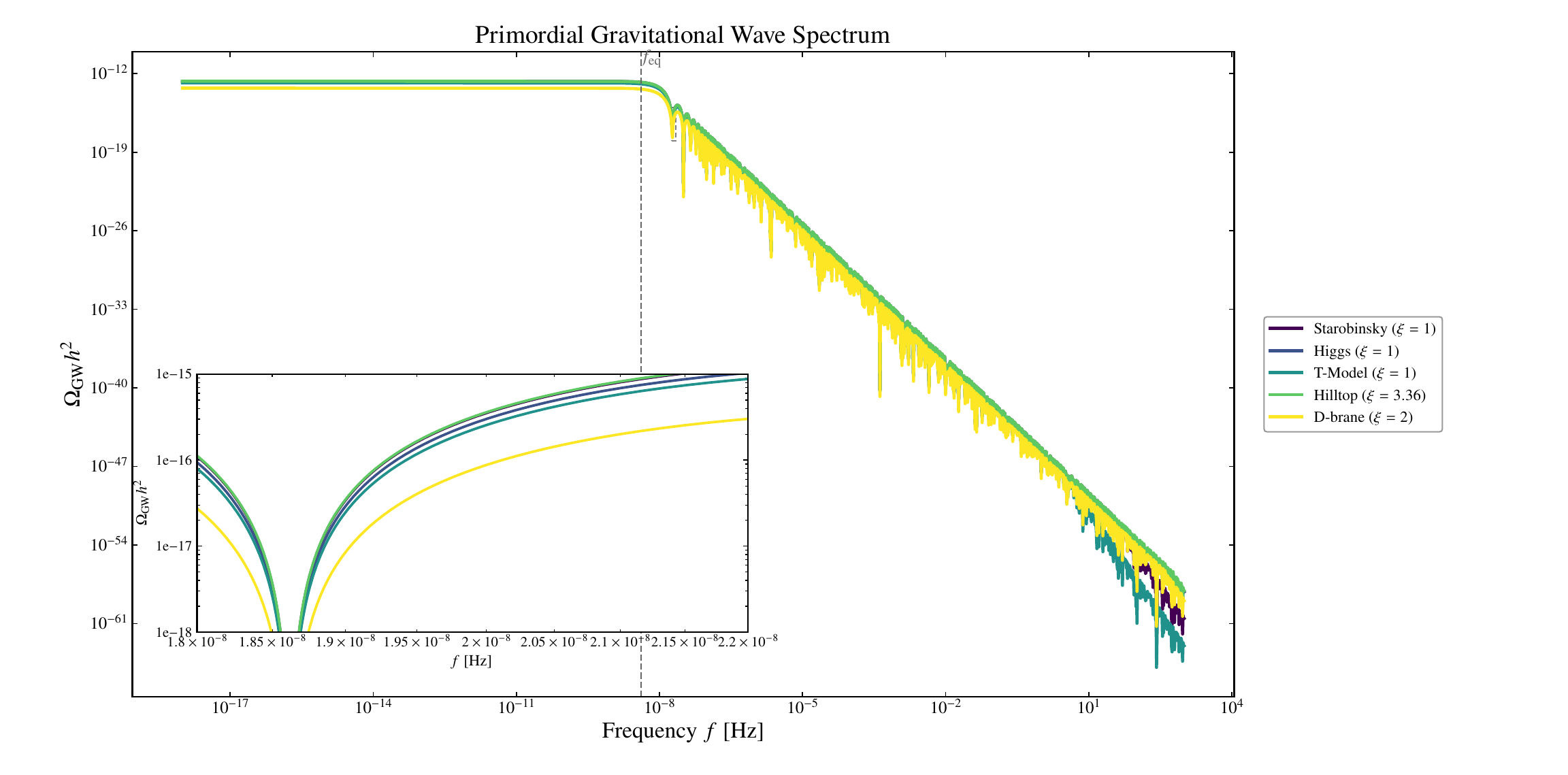}
    \caption{Primordial gravitational wave energy density spectrum, $\Omega_{\mathrm{GW}} h^{2}$, as a function of frequency $f$ for several inflationary models characterized by distinct tensor-to-scalar ratios $r$ and non-minimal coupling parameters $\xi$. The spectrum incorporates suppression due to reheating, matter--radiation equality, and neutrino damping. The vertical dashed line marks $f_{\mathrm{eq}}$, the frequency corresponding to horizon reentry at matter-radiation equality.}
    \label{fig:gw-spectrum}
\end{figure*}

Primordial gravitational waves (PGWs) provide a pristine observational window into the high-energy physics of inflation, unaltered by microphysical interactions at later times. The amplitude of the primordial tensor power spectrum directly probes the inflationary Hubble scale via $H \sim \sqrt{r} M_{\rm Pl}$, where $r$ is the tensor-to-scalar ratio, making gravitational wave detection a crucial test of inflationary physics \cite{Liddle_1993, Baumann:2009ds}.

The dimensionless energy density in GWs per logarithmic frequency interval today is given by
\begin{equation}
\Omega_{\rm GW}(k) = \frac{1}{12} \left( \frac{k}{a_0 H_0} \right)^2 \Delta_t^2(k)\, \mathcal{T}^2(k),
\end{equation}
where $\Delta_t^2(k)$ is the primordial tensor power spectrum and $\mathcal{T}(k)$ is the transfer function encoding subhorizon evolution through the various cosmological epochs. This expression captures the fact that $\Omega_{\rm GW}(f)$ is sensitive not only to the inflationary amplitude but also to the entire thermal history of the Universe following inflation, including reheating, changes in relativistic degrees of freedom, and neutrino free-streaming \cite{Watanabe:2006qe}.

\subsection{GW Spectrum and non-minimal coupling}

Figure~\ref{fig:gw-spectrum} illustrates $\Omega_{\rm GW} h^2(f)$ for a range of inflationary scenarios characterized by different values of the tensor-to-scalar ratio $r$ and the non-minimal coupling parameter $\xi$ between the inflaton and gravity. In models with $\xi \gg 1$, such as Higgs inflation or $R^2$ (Starobinsky) inflation, the inflationary potential becomes exponentially flat in the Einstein frame, strongly suppressing $r$ and consequently the gravitational wave amplitude \cite{Bezrukov:2007ep, Kallosh:2013hoa}. These scenarios predict an extremely faint $\Omega_{\rm GW}(f)$, often below the reach of upcoming observatories.

Conversely, models with small $\xi$ or those belonging to the class of $\alpha$-attractor potentials---particularly pole-type T-models---exhibit enhanced tensor amplitudes for a given number of $e$-folds, and hence higher $\Omega_{\rm GW}(f)$ values at CMB and interferometer-accessible frequencies \cite{Kallosh:2013hoa, Bartolo:2016ami}. These features make such models compelling targets for space-based GW missions like LISA, DECIGO, BBO, and terrestrial atomic interferometers like MAGIS and AEDGE \cite{Canuel:2019abg, Yagi:2011wg}.

\subsection{Suppression Mechanisms and Transfer Functions}

A prominent feature in the GW spectrum is the suppression near the frequency $f_{\rm eq} \simeq 10^{-16}\,\mathrm{Hz}$ corresponding to matter-radiation equality. The transition in expansion rate at this epoch alters the subhorizon evolution of tensor modes, resulting in a broken spectrum described approximately by:
\begin{equation}
\mathcal{T}^2(k) \simeq
\begin{cases}
1, & k \ll k_{\rm eq}, \\
\left( \frac{k_{\rm eq}}{k} \right)^2, & k \gg k_{\rm eq},
\end{cases}
\end{equation}
reflecting the decay in tensor mode energy for modes re-entering the horizon during matter domination \cite{Turner:1993vb}.

Additionally, the free-streaming of relativistic neutrinos introduces anisotropic stress that damps tensor perturbations on subhorizon scales, reducing $\Omega_{\rm GW}$ by up to $\sim 35\%$ at high frequencies \cite{Weinberg:2003ur}.

\subsection{CMB B-modes and Tensor Constraints}

The most stringent constraints on the inflationary tensor amplitude arise from B-mode polarization of the Cosmic Microwave Background (CMB), where tensor perturbations source a unique curl pattern. The B-mode signal peaks near multipoles $\ell \sim 80$, corresponding to $k \sim 0.01\, \mathrm{Mpc}^{-1}$, or a frequency:
\begin{equation}
f = \frac{k}{2\pi a_0} \approx 1.5 \times 10^{-17}\, \mathrm{Hz} \left( \frac{k}{0.01\, \mathrm{Mpc}^{-1}} \right),
\end{equation}
which lies at the lower end of the GW spectrum accessible via indirect observations.

The angular power spectrum of the B-mode polarization scales approximately as:
\begin{equation}
C_\ell^{BB} \propto r\, \Delta_\zeta^2(k)\, T_\ell^2,
\end{equation}
where $\Delta_\zeta^2(k)$ is the scalar power spectrum amplitude and $T_\ell$ is the tensor transfer function, sharply peaked at horizon crossing near recombination. In plateau-like slow-roll potentials such as Starobinsky or Higgs inflation, the predicted tensor-to-scalar ratio is $r \sim (2.5$--$4) \times 10^{-3}$, yielding $C_\ell^{BB} \sim 10^{-2}\, \mu\mathrm{K}^2$ at the recombination bump \cite{Kallosh:2013hoa}. These signals are within the projected sensitivity of LiteBIRD \cite{LiteBIRD:2022cnt}, CMB-S4 \cite{Abazajian:2016yjj}, and the Simons Observatory \cite{SimonsObservatory:2018koc}, which aim to reach $\sigma(r) \sim 10^{-4}$ levels.

\subsection{Reheating Imprints on the GW Spectrum}

The relic gravitational-wave (GW) spectrum spans a wide range of frequencies, from $f \sim 10^{-18}$ Hz up to GHz scales, corresponding to tensor modes re-entering the horizon across different cosmic epochs. The present-day spectrum can be approximately expressed as:
\begin{equation}
\Omega_{\rm GW}(f)\, h^2 \approx \frac{r}{24} \Delta_\zeta^2(k_*) \left( \frac{f}{f_*} \right)^{n_t} \mathcal{T}^2(f),
\end{equation}
where $f_* \simeq 1.5 \times 10^{-17}$ Hz is the CMB pivot frequency, and $n_t \approx -r/8$ under the slow-roll consistency relation. This relation may, however, receive scale-dependent corrections due to higher-order slow-roll effects, particularly when extrapolated over wide frequency baselines.

Throughout this work, the quoted values of $T_{\rm reh}$ correspond to benchmark estimates obtained in the instantaneous-reheating limit. In a more general scenario where reheating is prolonged and characterized by an effective equation-of-state parameter $w_{\rm reh}$, both the characteristic turnover frequency and the high-frequency slope of the GW spectrum would be modified. The quoted values should therefore be interpreted as reference scales for model comparison rather than precise determinations of the reheating history.

During reheating, the expansion of the Universe is governed by an effective fluid with equation-of-state parameter $w_{\rm reh}$. Tensor modes re-entering the horizon during this epoch evolve according to
\begin{equation}
\Omega_{\rm GW}(f) \propto f^{\frac{2(3w_{\rm reh}-1)}{3w_{\rm reh}+1}},
\end{equation}
implying a blue-tilted spectrum for $w_{\rm reh} > 1/3$, which enhances the GW amplitude at high frequencies ($f \gtrsim 10^{-2}$ Hz). Such scenarios are therefore promising targets for future space-based and atom-interferometer GW detectors. Conversely, extended reheating phases with $w_{\rm reh} < 1/3$ lead to a suppression of high-frequency power, potentially rendering the signal undetectable even for moderately large values of $r$ \cite{Kuroyanagi:2014nba}.

The reheating temperature $T_{\rm reh}$ further defines a characteristic transition frequency,
\begin{equation}
f_{\rm reh} \sim 0.1\, \mathrm{Hz} \left( \frac{T_{\rm reh}}{10^7\, \mathrm{GeV}} \right),
\end{equation}
above which the spectral slope becomes sensitive to the reheating equation of state and its duration. As a result, measurements of the GW spectrum at high frequencies provide a complementary probe of the post-inflationary expansion history and can help break degeneracies between inflationary models that are otherwise indistinguishable at CMB scales.

\subsection{Current Observational Constraints}

Observational bounds on the stochastic GW background currently include:
\begin{itemize}
    \item $r < 0.036$ (95\% CL), from joint BICEP/Keck and \textit{Planck} polarization data \cite{BICEP:2021xfz, Tristram:2021tvh},
    \item $\Omega_{\rm GW}(f \sim 100\,\mathrm{Hz})\, h^2 < 6 \times 10^{-9}$ from LIGO-Virgo O3 data \cite{LIGOScientific:2021nrg},
    \item A potential common-spectrum stochastic signal at $f \sim 10^{-9}$ Hz, observed consistently across pulsar timing arrays including NANOGrav \cite{NANOGrav:2023hvm}, EPTA \cite{EPTA:2023fyk}, and anticipated by SKA projections \cite{Janssen:2014dka}.
\end{itemize}

\subsection{Implications for Inflaton Dynamics}

A crucial theoretical consistency relation arises from the Lyth bound \cite{Lyth:1996im}, linking the tensor amplitude $r$ to the total field displacement $\Delta \phi$ of the inflaton over the last $N$ $e$-folds of inflation:
\begin{equation}
\Delta \phi \gtrsim \frac{M_{\rm Pl}}{\sqrt{8}} \left( \frac{r}{0.01} \right)^{1/2} N.
\end{equation}
This inequality implies that large $r$ values---and hence detectable $\Omega_{\rm GW}$ signals---require super-Planckian field excursions ($\Delta\phi > M_{\rm Pl}$ for $r \gtrsim 0.01$, $N = 50$--$60$), characteristic of large-field models like chaotic inflation or $\alpha$-attractors. In contrast, small-field models such as hybrid or hilltop inflation involve sub-Planckian excursions, leading to suppressed GW signatures. Nevertheless, such models remain within reach of ultra-sensitive missions like LiteBIRD and CMB-S4.

Taken together, the primordial GW spectrum encodes an extraordinary breadth of information about both the inflationary dynamics and the post-inflationary thermal history of the universe. The combination of B-mode polarization, space-based interferometry, and PTA observations offers a multi-band strategy to probe inflationary energy scales, test the Lyth bound, and constrain the nature and duration of reheating.

\section{Model Comparison and Predictions}
\label{sec:model_comparison}

Table~\ref{tab:GW_Treh_summary} summarizes the inflationary predictions and reheating features of several theoretically motivated benchmark models, evaluated consistently at $N_e = 60$ $e$-folds. The observables compared include the scalar spectral index $n_s$, tensor-to-scalar ratio $r$, inflaton field excursion $\Delta\phi$, inflationary Hubble scale $H_{\rm inf}$, and the reheating temperature $T_{\rm reh}$. Together, these quantities characterize not only the primordial scalar and tensor fluctuations, but also the post-inflationary thermalization history---crucial for interpreting observational data from CMB polarization and gravitational wave (GW) detectors. Crucially, the $T_{\rm reh}$ column should be read as an instantaneous-reheating benchmark. If reheating lasts for a finite duration and is characterized by an effective equation-of-state parameter $w_{\rm reh}$, the inferred $T_{\rm reh}$, the mapping between $N_e$ and the CMB pivot scale, and the high-frequency branch of $\Omega_{\rm GW}(f)$ can all shift. The comparative conclusions based on $(n_s,r,\alpha_s,\beta_s)$ are comparatively robust, while statements involving $T_{\rm reh}$ and high-frequency gravitational waves are correspondingly more model-dependent.

\vspace{1em}
\noindent\textbf{Starobinsky Inflation (Plateau Model):}  
As a prototype of $R + R^2$-type potentials, Starobinsky inflation \cite{Starobinsky:1980te} predicts a highly suppressed tensor amplitude, $r \approx 3.8 \times 10^{-3}$, and a spectral index $n_s \approx 0.963$, well within the $1\sigma$ region favored by current CMB data \cite{Planck:2018jri}. The inflaton field excursion remains sub-Planckian, $\Delta\phi \lesssim 1\,M_{\rm Pl}$, and the energy scale of inflation is $H_{\rm inf} \sim 1.8 \times 10^{14}\,\mathrm{GeV}$. Within the instantaneous-reheating benchmark adopted here, the model corresponds to a high $T_{\rm reh} \sim 2.0 \times 10^{14}$ GeV; a detailed entropy-production history would require specifying the microscopic reheating channel.

\vspace{1em}
\noindent\textbf{Higgs Inflation (non-minimal coupling):}  
Higgs inflation, driven by the Standard Model Higgs field with a non-minimal coupling $\xi \gg 1$ to gravity, transitions to a plateau potential in the Einstein frame \cite{Bezrukov:2007ep}. At small $\xi$, the predicted $r$ can reach $\sim 6.4 \times 10^{-2}$, decreasing rapidly as $\xi$ increases, with $r \sim 3.5 \times 10^{-3}$ for $\xi \sim 1$. The inflationary scale decreases correspondingly from $H_{\rm inf} \sim 6.65 \times 10^{13}$ to $1.55 \times 10^{13}$ GeV. Despite the relatively low energy scale, the reheating temperature remains high, $T_{\rm reh} \sim 10^{15}\,\mathrm{GeV}$, enabled by strong Yukawa and gauge couplings to Standard Model particles \cite{Bezrukov:2007ep}.

\vspace{1em}
\noindent\textbf{T-Models ($\alpha$-Attractors):}  
These models represent a broad class of inflationary attractors where the potential exhibits a pole structure near $\phi \to 0$ in the Einstein frame \cite{Kallosh:2013hoa}. The parameter $\alpha$ governs the curvature of the scalar field manifold and interpolates between large-field ($\alpha \gg 1$) and plateau ($\alpha \ll 1$) behavior. For large $\alpha$, tensor amplitudes approach $r \sim 0.1$, consistent with chaotic inflation limits, while small $\alpha$ yields $r \ll 10^{-4}$, placing such models at the edge of detectability by LiteBIRD or CMB-S4 \cite{LiteBIRD:2022cnt, Abazajian:2016yjj}. The corresponding $H_{\rm inf}$ values range from $7.6 \times 10^{15}$ GeV down to $7.5 \times 10^{14}$ GeV. Reheating in these models is typically inefficient due to weak model-dependent couplings to visible matter, leading to $T_{\rm reh} \sim 10^{11}$--$10^{13}$ GeV.

\vspace{1em}
\noindent\textbf{Hilltop Quartic Inflation:}  
This class of small-field models posits inflation near a local maximum, yielding moderately small tensor amplitudes ($r \sim 10^{-3}$) and $\Delta\phi \sim 0.4\,M_{\rm Pl}$ \cite{Martin:2013tda}. Although the potential curvature near the hilltop enhances scalar modes, the tensor sector remains suppressed. The inflationary scale is reduced to $H_{\rm inf} \sim 1.5$--$1.7 \times 10^{13}$ GeV, but reheating can be moderately efficient with $T_{\rm reh} \sim 3.2$--$3.5 \times 10^{15}$ GeV, depending on the coupling structure to reheating channels.

\vspace{1em}
\noindent\textbf{D-brane Inflation (String-Theoretic):}  
Originating from warped flux compactifications in type IIB string theory, D-brane inflation involves the motion of a brane--antibrane pair in extra-dimensional throats \cite{Kachru:2003sx, Baumann:2006th}. These models generically predict ultra-small tensor modes, $r \sim 10^{-3}$ or lower, and low $H_{\rm inf} \sim 8 \times 10^{12}$ GeV due to steep potentials and backreaction. Nevertheless, reheating can be highly efficient owing to closed-string radiation and brane annihilation, with $T_{\rm reh} \sim 2.4 \times 10^{15}$ GeV.

\vspace{1em}
These results underscore the power of gravitational wave observables in discriminating among inflationary scenarios that appear degenerate in the $(n_s, r)$ plane. For example, Starobinsky and $\alpha$-attractors with small $\alpha$ predict identical $(n_s, r)$ values but diverge significantly in reheating temperature and GW spectral shape in the mid-frequency regime ($f \sim 10^{-2}$--$1$ Hz). These frequencies are within reach of next-generation interferometers such as DECIGO and BBO \cite{Yagi:2011wg}, and space-based detectors like LISA \cite{LISA:2017pwj}.

The redshifted tensor spectrum, $\Omega_{\rm GW}(f) h^2$, shown in Figure~\ref{fig:gw-spectrum}, encapsulates these differences. All models converge to a scale-invariant plateau at high frequencies, modulated by the tensor tilt $n_t$ and reheating-induced suppression at characteristic scales $f \sim f_{\rm reh}$. In particular, Hilltop and T-models with moderately high $r$ values exhibit elevated $\Omega_{\rm GW}(f)$ in the $f \sim 0.01$--$1$ Hz regime, approaching sensitivity limits of planned missions. D-brane scenarios, while exhibiting suppressed tensor spectra at CMB scales, can produce comparatively large $\Omega_{\rm GW}$ at higher frequencies due to exotic preheating or brane annihilation dynamics.

Furthermore, degeneracies in $(n_s, r)$ can be broken via joint constraints on $(T_{\rm reh}, H_{\rm inf}, \Delta\phi)$ and the frequency-dependent structure of $\Omega_{\rm GW}(f)$. Models that share the same CMB-scale tensor amplitude can differ by several orders of magnitude in GW amplitude at $f \sim 10^{-1}$ Hz depending on their post-inflationary equation of state and thermalization mechanisms \cite{Kuroyanagi:2014nba}.

This emphasizes the critical synergy between CMB polarization (LiteBIRD, CMB-S4), pulsar timing arrays (NANOGrav, SKA), and direct GW probes (LISA, BBO, AEDGE). A coordinated multi-frequency reconstruction of $\Omega_{\rm GW}(f)$ promises to chart the inflationary potential, the reheating phase, and the underlying field-theoretic or string-theoretic origins of the early universe.

\section{Conclusion}
\label{sec:conc}

In this study, we have conducted a detailed numerical analysis of five theoretically well-motivated single-field inflationary models with non-minimal gravitational couplings---namely, Starobinsky inflation, Higgs inflation, T-models within the $\alpha$-attractor framework, quartic Hilltop inflation, and D-brane inflation. By leveraging updated observational bounds from \textit{Planck}, ACT DR6, and DESI DR2, alongside projected sensitivities from upcoming probes such as LiteBIRD, CMB-S4, and space-based gravitational wave interferometers, we have systematically mapped out the observational signatures that can differentiate these models across both the inflationary and post-inflationary epochs. The interpretation of model preference remains dataset-dependent: adopting the SPT/Planck central value keeps Starobinsky-like and Higgs-like plateau predictions close to the preferred region, whereas adopting the P-ACT-LB central value shifts these models toward the lower-$n_s$ side and makes larger-$n_s$ scenarios, including certain hilltop realizations, comparatively more competitive.

\textbf{Starobinsky Inflation} continues to exemplify the observational gold standard of plateau models, with theoretical predictions of $n_s \approx 0.963$, $r \sim 3.8 \times 10^{-3}$, and controlled sub-Planckian field excursions $\Delta\phi \lesssim 0.4\, M_{\rm Pl}$. Its predictions remain firmly within current and anticipated CMB $1\sigma$ contours, even when accounting for the running parameters $(\alpha_s, \beta_s)$ and variations in the reheating temperature $T_{\rm reh}$. Owing to its geometric origin in $R + R^2$ gravity \cite{Starobinsky:1980te}, the model is both UV-insensitive and highly predictive.

\textbf{Higgs Inflation}, constructed via a large non-minimal coupling $\xi$ to the Ricci scalar, exhibits stronger parameter sensitivity. As $\xi$ increases, the model flows toward Starobinsky-like behavior, suppressing $r$ and bringing $\Delta\phi$ into the sub-Planckian regime \cite{Bezrukov:2007ep,Bezrukov:2013fka}. However, the inflationary scale $H_{\rm inf}$ and the corresponding reheating temperature $T_{\rm reh}$ exhibit nonlinear dependence on $\xi$, introducing observable variations in the tensor tilt and higher-order spectral parameters. Models with $\xi \gtrsim 0.1$ remain viable under current bounds and yield $T_{\rm reh} \sim 10^{15}\, \mathrm{GeV}$ due to efficient coupling to the Standard Model sector.

\textbf{T-Models}, arising from $\alpha$-attractor scenarios in supergravity and conformal field theory \cite{Kallosh:2013hoa,Galante:2014ifa}, interpolate between large-field and plateau limits, with tensor-to-scalar ratios spanning $r \sim 10^{-6}$ to $r \sim 10^{-1}$ as $\alpha$ is varied. For small $\alpha \lesssim 1$, these models predict ultra-suppressed $r$, enhanced $\beta_s \sim \mathcal{O}(0.1)$, and low-scale inflation---distinguishing them in the $(\alpha_s, \beta_s)$ plane despite their degeneracy with plateau models in the $(n_s, r)$ plane. While these predictions remain below current detection thresholds, upcoming surveys such as SPHEREx and CMB-HD may provide the necessary sensitivity.

\textbf{Hilltop Quartic Inflation}, characterized by small-field evolution near a local maximum, exhibits moderate tensor signatures ($r \sim 10^{-3}$) and short field excursions $\Delta \phi \sim 0.4\, M_{\rm Pl}$. Though it is marginally consistent with current constraints at $N = 60$, it is increasingly disfavored by future forecast ellipses in the $(n_s, \alpha_s)$ plane. Remarkably, it predicts the highest amplitude primordial GW background in the intermediate $f \sim 10^{-2}$--$1$ Hz regime, with $\Omega_{\rm GW}(f)$ approaching the sensitivity limits of DECIGO, BBO, and MAGIS \cite{Kuroyanagi:2014nba,Yagi:2011wg}. This positions Hilltop inflation as a strong candidate for direct GW detection, especially when accounting for preheating-induced amplification effects.

\textbf{D-brane Inflation}, embedded in string theory compactifications with warped throat geometries \cite{Kachru:2003sx,Baumann:2006cd}, predicts low tensor amplitudes ($r \sim 10^{-3}$), sub-Planckian excursions ($\Delta\phi \sim 0.19\, M_{\rm Pl}$), and stable $n_s \approx 0.963$ over a broad range of $\xi$. While its spectral index slightly overshoots forecasted $1\sigma$ regions, its string-theoretic robustness and reheating via brane-antibrane annihilation ($T_{\rm reh} \sim 10^{15}\, \mathrm{GeV}$) render it a theoretically compelling model, particularly in contexts where UV consistency is prioritized.

Overall, our analysis demonstrates that a simple $(n_s, r)$ portrait is insufficient to discriminate between models in the precision cosmology era. Instead, the inclusion of higher-order parameters $(\alpha_s, \beta_s)$, the tensor tilt $n_t$, reheating dynamics ($T_{\rm reh}$, $w_{\rm reh}$), and the full spectral structure of $\Omega_{\rm GW}(f)$ provide a richer model-selection framework. In particular, multi-frequency GW data---from ultra-low frequency CMB $B$-modes ($f \sim 10^{-17}$ Hz), intermediate-band space interferometers (mHz--Hz), and pulsar timing arrays ($f \sim 10^{-9}$ Hz)---form a crucial triad for disentangling inflationary scenarios with otherwise degenerate predictions.

Even if, in the near future, primordial gravitational waves remain undetectable through the tensor-to-scalar ratio $r$, the accumulation of precise statistical evidence from the $(\alpha_s, n_s)$ parameter space across diverse cosmological observations can still robustly constrain---and in some cases rule out---broad classes of inflationary models, including those analyzed in this paper.

A key discriminator is the Lyth bound \cite{Lyth:1996im}, which links field excursions to tensor amplitudes. Large-field models with $\Delta \phi \gtrsim M_{\rm Pl}$ yield detectable GWs, while small-field models remain sub-luminous in this channel. Thus, the field-range--tensor-amplitude duality becomes central in separating UV-complete string models from effective field theories with minimal excursions.

The references cited in this forward-looking paragraph are included only to identify the broader PBH and stochastic-inflation literature motivating future work. They are not used in constructing the likelihood comparison, the numerical slow-roll pipeline, the reheating benchmark estimates, or the gravitational-wave spectra reported in this work.

As a next step, we plan to investigate the feasibility of primordial black hole (PBH) formation in these single-field models. Given recent claims that ACT DR6 data disfavors PBH scenarios \cite{Frolovsky:2025iao}, a re-evaluation of power spectrum enhancement, non-Gaussian statistics, and tail behavior is warranted. Following recent developments \cite{Choudhury:2023vuj,Choudhury:2023jlt,Choudhury:2023rks,Choudhury:2023hvf,Choudhury:2023kdb,Choudhury:2023hfm,Bhattacharya:2023ysp,Choudhury:2023fwk,Choudhury:2023fjs,Choudhury:2024one,Choudhury:2024ybk,Choudhury:2024jlz,Choudhury:2024dei,Choudhury:2024dzw,Choudhury:2024aji,Choudhury:2024kjj}, we aim to construct a data-driven, nonperturbative framework for PBH formation that respects current bounds while illuminating regions of parameter space consistent with early-universe gravitational instability.

In parallel, a promising and timely direction is to confront recently derived next-to-next-to-next-to-leading-order (N3LO) predictions for Starobinsky inflation with current and upcoming observations. These N3LO results, expressed analytically and in a frame-invariant form in terms of the scalar tilt $n_s$, offer a new precision frontier for inflationary theory~\cite{bianchi2025precisionpredictionsstarobinskyinflation,Bianchi_2024}. Leveraging our existing analysis of $(n_s, \alpha_s, \beta_s)$ constraints from ACT DR6 and DESI DR2, we aim to test the compatibility of these high-order predictions with data and explore whether the running and running of the running of the scalar spectral index can provide empirical support for or against the Starobinsky paradigm. Such a comparison represents a critical benchmark for precision cosmology and an opportunity to link observational cosmology directly with UV-complete inflationary dynamics.

In summary, the inflationary landscape is entering a phase of empirical scrutiny across multiple cosmological windows. The joint use of spectral observables, gravitational waves, reheating diagnostics, and now analytic N3LO predictions enables a powerful and multidimensional discriminator among inflationary theories---one that is rapidly becoming a necessary tool to chart the universe's earliest moments with precision.

\section*{Acknowledgement}
SC would like to thank Md. Sami for the useful discussions. Research of GB is funded by the Science Committee of the Ministry of Science and Higher Education of the Republic of Kazakhstan (Grant AP19175860).

\bibliographystyle{utphys}

\end{document}